\documentclass[sigconf,natbib=false]{acmart}

\title{Law and the Emerging Political Economy of Algorithmic Audits}

\hypersetup{
breaklinks=true,   
colorlinks=true,   
}

\usepackage[utf8]{inputenc}

\RequirePackage[
  datamodel=acmdatamodel,
  style=acmnumeric,backend=biber
  ]{biblatex}
  
\author{Petros Terzis}
\email{p.terzis@uva.nl}
\orcid{0000-0001-8985-5651}
\affiliation{%
  \institution{University of Amsterdam}
  \department{Institute for Information Law}
  \city{Amsterdam}
  \country{the Netherlands}
}
\additionalaffiliation{%
  \institution{University College London}
  \department{Faculty of Laws}
  \city{London}
  \country{United Kingdom}
}

\author{Michael Veale}
\email{m.veale@ucl.ac.uk}
\orcid{0000-0002-2342-8785}
\affiliation{%
  \institution{University College London}
  \department{Faculty of Laws}
  \city{London}
  \country{United Kingdom}
}
\additionalaffiliation{%
  \institution{University of Amsterdam}
  \department{Institute for Information Law}
  \city{Amsterdam}
  \country{The Netherlands}
}

\author{No\"{e}lle Gaumann}
\email{noelle.gaumann.20@ucl.ac.uk}
\affiliation{%
  \institution{University College London}
  \department{Faculty of Laws}
  \city{London}
  \country{United Kingdom}
}

\keywords{auditing, algorithmic audits, Digital Services Act, Online Safety Act, political economy}

\ccsdesc[500]{Social and professional topics~Computing / technology policy}
\ccsdesc[500]{Information systems~Social networking sites}

\setcopyright{rightsretained}
\acmYear{2024}
\acmConference[ACM FAccT '24]{the 2024 ACM Conference on Fairness, Accountability, and Transparency}{3--6 June 2024}{Rio de Janeiro, Brazil}

\begin{abstract}
	For almost a decade now, scholarship in and beyond the ACM FAccT community has been focusing on novel and innovative ways and methodologies to audit the functioning of algorithmic systems. Over the years, this research idea and technical project has matured enough to become a regulatory mandate. Today, the Digital Services Act (DSA) and the Online Safety Act (OSA) have established the framework within which technology corporations and (traditional) auditors will develop the `practice’ of algorithmic auditing thereby presaging how this `ecosystem’ will develop. In this paper, we systematically review the auditing provisions in the DSA and the OSA in light of observations from the emerging industry of algorithmic auditing. Who is likely to occupy this space? What are some political and ethical tensions that are likely to arise? How are the mandates of `independent auditing’ or `the evaluation of the societal context of an algorithmic function’ likely to play out in practice? By shaping the picture of the emerging political economy of algorithmic auditing, we draw attention to strategies and cultures of traditional auditors that risk eroding important regulatory pillars of the DSA and the OSA. Importantly, we warn that ambitious research ideas and technical projects of/for algorithmic auditing may end up crashed by the standardising grip of traditional auditors and/or diluted within a complex web of (sub-)contractual arrangements, diverse portfolios, and tight timelines.
\end{abstract}  

\begin{document}

\maketitle

\section{Introduction}\label{introduction}

The idea of auditing online platforms and algorithms as a form of governance is not new nor is its critique \autocite{brownAlgorithmAuditScoring2021,costanzachockWhoAuditsAuditors2022,imanaHavingYourPrivacy2023,rajiClosingAIAccountability2020,rajiAlgorithmicAuditsActual2022}. From the drafting of high-level ethical principles and codes of conduct to the technical development of `fairness toolkits' and standards, scholarship in FAccT and beyond has been focusing on ways to translate normative mandates into actionable corporate practices. Today however, regulations have incorporated forms of these efforts, and we see early signs of the institutional and organisational dynamics that are likely to shape the future trajectories of algorithmic governance.

Looking systematically at the various legal developments in the field of algorithmic auditing in Global North jurisdictions suggests that auditing and certification assessment as a practice of entrusting to third-parties the evaluation of certain properties of AI and IT systems --- reliability, security, fairness, transparency and privacy and more --- is here to stay. In the US, there have been repeated attempts to introduce federal regulation recommending independent audits for meaningful human rights impact assessments. Proposals have been made at the federal level \autocite{AlgorithmicAccountabilityAct2023}, in local legislation in New York City \autocites{LocalLawInt2023}{grovesAuditingWorkExploring2024}, and the Federal Trade Commission (FTC) has expanded on some of the fundamental principles that such audits shall follow \autocites[50--58]{federaltradecommissionCombattingOnlineHarms2022}{ramdasIdentifyingActionableAlgorithmic2022}{senwydenronAlgorithmicAccountabilityAct2022}. Canada and Australia are also set to introduce obligations for internal and external audits as accountability mechanisms for various algorithmic systems \autocites{goodmanAIAuditWashingAccountability2022}{australiangovernmentRoyalCommissionRobodebt2023}{governmentofcanadaSummaryRegulatoryPowers2022}. In the European Union, this trend is particularly strong. The AI Act introduces a framework for conformity assessment which, for some applications, requires a `notified body' to audit documentation of compliance  \autocites{IntroductionConformityAssessment2022}{vealeDemystifyingDraftEU2021}. In the same spirit, under art. 15 of the Digital Markets Act (DMA), entities designated as `gatekeepers' will have to undergo an independent audit for any techniques of consumers' profiling they deploy, whereas under art. 23(3) of the same, the Commission will be able to appoint external auditors as part of its power to conduct inspections. Finally, indicative of the popularity of audit-like mechanisms for technology regulation is the fact that the European Parliament has suggested an AI Act-like conformity assessment procedure for the first (of many) sector-specific regulation for European `data spaces', namely the proposed European Health Data Space (EHDS) \autocite{europeanparliamentEuropeanHealthData2023,terzisInteroperabilityGovernanceEuropean2023}.

This complicated regulatory canvas risks blurring the lines of what is expected to be auditable and audited, at which stage of the development, by whom, and to what end \autocite{rajiClosingAIAccountability2020}. In this frenzy of legal developments, concepts such as risk-assessment and conformity assessment, or compliance and auditing, are often thrown around seemingly interchangeably \autocite{goodmanALGORITHMICAUDITINGCHASING2023}. Absent clarity, responsibility for auditing disperses amongst various actors, internal and external forms of control are becoming difficult to differentiate and evaluate, benchmarks and standards vary, and accountability is diluted across a complicated value chain of/for compliance \autocite{cobbeUnderstandingAccountabilityAlgorithmic2023}. However, this does not by itself mean that law(s) will not work. On the contrary, it suggests that with legal regimes now in place, the decision-making authority for resolving tensions and inconsistencies is departing from the realm of politics and/or research and migrates to that of law by recognising certain actors as legitimate bearers of the responsibility for/of audit.

To achieve further clarity, this paper systematically reviews the provisions of the UK's Online Safety Act (OSA) 2023, the EU's Digital Services Act (DSA) 2022 and the latter's associated Delegated Regulation on the performance of audits (DRPA) 2023, which taken together arguably present the most comprehensive proposed framework for understanding the dynamics of external auditing of large online platforms and their algorithmic systems. We start from the letter of the law, but do not end there. Instead, through an exploration of consultation submissions of corporate actors, their advertised portfolios and projects, and the industry dynamics in the field of algorithmic auditing, the paper situates the legal analysis within the emerging political economy of algorithmic auditing: Who is likely to occupy this space? What are some political and ethical tensions that are likely to arise? How are the legal mandates of independent auditing likely to play out in practice? Section \ref{comparative-analysis-of-the-online-safety-act-and-the-digital-services-act} is descriptive in that it sets out, in a systematic way, the main provisions that will guide the shaping of the algorithmic auditing ecosystem in the EU and the UK moving forward. Section \ref{potential-trajectories} draws on this analysis and observes tensions, challenges, and dilemmas that are likely to arise in practice. And, eventually, Section \ref{elements-of-the-emerging-political-economy-of-algorithmic-auditing} provides a roadmap of likely scenarios for the future of algorithmic auditing and the uncomfortable reality that researchers in this space might ultimately have to confront and grapple with.

\section{Comparative Analysis of the Online Safety Act and the Digital Services Act}\label{comparative-analysis-of-the-online-safety-act-and-the-digital-services-act}

\subsection{Overview}\label{overview}

Throughout their short history, many social media platforms developed their own systems for controlling, monitoring and moderating the flow of content on their platforms, often as a result of controversy, and with a strong role for algorithmic systems \autocite{gorwaPoliticsPlatformRegulation2024}. Following a `grace' period of self-regulation that lasted more than a decade, the rule-making and rule-monitoring dynamics underpinning the functioning of social media platforms look now to change. In this context, the UK's OSA and the EU's DSA have been touted as solutions. They are ambitious instruments that are expected to enable governmental oversight over crucial issues such as the spread of hateful, misinforming, or illegal content, operating by removing aspects of decision-making authority away from private entities and by establishing formal guidelines and processes. Both pieces of legislation can arguably be understood as a by-product of our changed understanding of online platforms and in particular social media platforms as transnational public forums for all matters social, political and otherwise \autocite{imanaHavingYourPrivacy2023,iosifidisPublicSphereSocial2011}. Both depart from the classic policy hope that rendering firms liable for illegal content upon notice of its existence provides enough incentive for a `healthier' internet \autocite{edwardsGreatPowerComes2019,macsithighRoadResponsibilitiesNew2020},\footnote{Note that the United States is a notable outlier in this, unusual globally in that platforms are immunised against liability for most types of illegal content other than IP infringement regardless of whether they have been given clear and specific notice of its existence. This is the famous `Section 230' (of the Communications Act, commonly misstated by almost everyone as Section 230 of the Communications Decency Act, where it is actually found in Section 9).}  towards direct responsibilities, typically framed as procedural and due diligence duties and obligations \autocite{buitenDigitalServicesAct2022}.

\subsubsection{Legislative Background: Online Safety Act}

The UK's Online Safety Act 2023 has its roots over 6 years prior in the creation of the 2017 Internet Safety Green Paper \autocite{internetsafetygreen}, itself following pledges around social media in the 2017 Conservative Party Manifesto, and poorly thought through (and thus never-commenced\footnote{Legislation can be passed, but have a provision that states that certain parts are not active until an order from the Secretary of State --- this is called commencement.}) legislation on website age verification in the Digital Economy Act 2017. Its turbulence has been connected to both the intrinsically contested nature of speech governance and the fact that there have been seven consecutive responsible Secretaries of State for the relevant portfolio (then Digital, Culture, Media and Sport), spanning four Prime Ministers, in the period between the Green Paper and Royal Assent of the Act.  The form of the act was initially inspired by the work of Professor Lorna Woods at the University of Essex and Will Perrin at the Carnegie UK Trust, who proposed a `duty of care' for social media platforms \autocite{woodsObligingPlatformsAccept2021}, although their proposed draft (with Maeve Walsh) was succinct compared to the final Act, comprising of only 20 sections compared to the passed statute's 241 sections (and 65 pages of 17 further detailed Schedules) \autocite{woodsDraftOnlineHarm2019}.

Political opposition to the provisions regulating content that was `legal but harmful' to adults in the Online Safety Bill as introduced led to their eventual removal. There are multiple reasons for this, but one interpretation is that it clashed with an emerging more divisive, populist form of politics, as such provisions (and their subjective notions of offence or harm) interfere with the perceived political potency of mobilising opinion against marginalised groups, including transgender individuals and migrant communities. The final Act focuses primarily on illegal content in relation to adults, and both illegal content and legal content that may be harmful to children on services which are likely to be accessed by children. However, it is very difficult to create a service that is not `likely to be accessed by children' without putting in place significant age detection systems --- either verification using identity, or `assurance', a set of highly speculative technologies designed to algorithmically infer individuals' ages from biometrics or service usage data.

The form of the OSA is complex and there are a huge array of provisions and powers --- space prohibits us providing a full overview here, least of all a critical one. For our purposes however, the core structure is that online platforms have to assess risks that their services, including their algorithmic systems, facilitate, and after undertaking this assessment, they have certain mitigation duties. The range of risks that need to be assessed and mitigated are significantly wider when platforms have or might have children as users. Differently sized platforms have different obligations, although at the time of writing no threshold or test has been published. This is one of the many parts of the OSA that either are awaiting an executive action (the making of regulations by the Secretary of State), or can be amended with limited parliamentary oversight by the executive. Faced with the complexity of its tasks, the regulator Ofcom --- the existing telecoms and media regulator --- is in the process of publishing literally thousands of pages of discussion and guidance for consultation.

\subsubsection{Legislative Background: Digital Services Act}

The Digital Services Act (DSA) is a Regulation --- a type of European law that does not require Member States to manually rewrite (`transpose') it into their national law --- passed in 2022, around two years after it was proposed by the European Commission.\footnote{While it is confusingly called an `Act', this is legally meaningless, and is just a name that the EU is branding recently large Regulations with in order to make them sound considerably sexier and more PR-friendly.} It responds in part to varying national laws concerning expression on platforms, including Germany's Network Enforcement Act 2017 (Netzwerkdurchsetzungsgesetz, or NetzDG); Austria's Communication Platforms Act; and France's later struck-down `Avia' Law (loi 2020-766). Indeed, in principle, the EU cannot create legislation such as the Digital Services Act --- harmonisation legislation created under the Treaty on the Functioning of European Union (TFEU) art. 114 --- without evidence of actual or potential fragmentation of law in Member States which could damage the internal market. The DSA, like much of EU law, has a stated aim in making compliance easier across borders by simplifying and homogenising the rulebook, while also introducing substantively new provisions into many jurisdictions which to date have had limited-to-no domestic platform regulation.

The DSA follows over a decade of European attempts at self- and co-regulation of online services \autocites{marsdenInternetCoRegulationEuropean2011}{brownRegulatingCodeGood2013}{buitenDigitalServicesAct2022}, which culminated in the 2016 EU \textit{Code of Conduct on Countering Illegal Hate Speech Online}, the 2017 and 2018 \textit{Communication} and \textit{Recommendation on Illegal Content Online} respectively, and the 2018 \textit{Code of Practice on Disinformation} \autocite{marsdenPlatformValuesDemocratic2020a}. The DSA in some ways is more timid than the OSA, replicating nearly verbatim the existing intermediary liability provisions from the e-Commerce Directive 2000, including the prohibition on laws, regulators or court orders obliging general monitoring of all content on a platform.\footnote{In contrast in the UK, this provision was never transposed from EU law into UK law and thus its post-Brexit status is unclear.} Note that the DSA does not endeavour to create new forms of liability for content --- these flow from existing national and Union law --- but instead focuses on the information flow and management, including the processes around how this content is promoted, contested and removed. It has some procedural provisions that apply to all online platforms, and then a further range of provisions that apply to only Very Large Online Platforms (VLOPs)/Search Engines (VLOSEs). Many, although not all, of the provisions that we consider to be closely connected to audit are only applicable to these larger entities.

The scope and regulatory remit of the two regulations vary. In particular, where the OSA aspires to make the Internet a `safer' place by primarily encompassing provisions for safety-by-design and the removal of illegal content online, with a strong emphasis on children's safety, the DSA introduces rules and mechanisms whose constitutive goal is to imbue accountability along with `greater democratic control and oversight over (\ldots) platforms' \autocite{EUDigitalServices}. In both regulations the burden of responsibility correlates to the size and scale of the platform. In this direction, both OSA's Category 1 providers as well as the DSA's VLOPs and VLOSEs are/will be determined by secondary legislation and (will) have increased duties and responsibilities. The auditing obligations will follow a similar pattern, increasing in `severity' and scope along with the scale and size of the platforms involved. However, both regulations have a long tail of platforms in-scope, with the impact assessment of the OSA indicating 25,000 firms will be subject to many of its obligations \autocite{hmgovernmentImpactAssessmentOnline2022}.

The following section sets out the different audit provisions attached to these regimes by looking at internal audits, new audit-like regimes, and external audits. The term `external' is to be understood quite broadly here as we depart from an established typology of audits (first-party, second-party, and third-party) purely to maintain consistency with the examined legal provisions, and to use them as a starting point to understand how law itself may shape these practices \autocites{costanzachockWhoAuditsAuditors2022}{messmerAuditingRecommenderSystems2023}.

\subsection{Internal audits}\label{internal-audits}

Both the OSA and the DSA oblige certain platforms to conduct internal risk assessments on their services. These can be understood in part as forms of internal algorithmic auditing, although the obligations may be wider than that.

Starting from the former, under the OSA there are four different processes of risk assessments that may or may not be applicable according to the nature and size of an online platform. In particular, all platforms must conduct a risk assessment for illegal content (s. 9). Where services are likely to be accessed by children --- a test, as mentioned, made very hard for significant platforms to not meet --- the respective platform must also conduct a so-called `children's risk assessment' (s. 12). Finally, the designated `Category 1' providers will need to perform a `user empowerment' risk assessment to evaluate the inclusion (`to the extent that it is proportionate to do so') of design features for increasing the control of users on the content delivered through the platform (s. 14). Platforms are instructed specifically to take into account `(in particular) algorithms used by the service' in all these risk assessments (ss. 9(5)(b), 12(6)(b), 14(5(c))).

DSA's internal audits seem to be more expansive in scope and remit, although apply to fewer platforms. VLOPs and VLOSEs are obliged to `diligently identify, analyse and assess any systemic risk' {[}emphasis added{]} stemming from the functioning of their systems and services (art. 34). In particular, recitals 80--83 set out the four main categories of systemic risks that VLOPs and VLOSEs shall take into account, namely: a) the dissemination of illegal content; b) actual or foreseeable impact on the exercise of fundamental rights; c) actual or foreseeable impact on democratic processes; and d) actual or foreseeable negative effect on other social and political rights (including public health or gender-based violence). Again, such risks clearly incorporate many automated and algorithmic systems which will need assessing, and therefore already blur the concept of a risk assessment with algorithmic audit.

Even though these audits are internal, there are many hooks for public participation, seen as both important but under-tooled in the existing algorithmic audit landscape \cite{ojewaleAIAccountabilityInfrastructure2024}. In many ways, the public participation hook is seems clearer than other regimes, such as the involvement `where appropriate' for related UK and EU data protection impact assessments under the GDPR \autocite{christofiDataProtectionControl2022}. The DSA invites platforms to engage with groups affected by systemic risks (recital 90), and in a similar context, Ofcom recommends consulting a variety of stakeholders including users and experts when conducting risk assessments, which may also include commissioned research \autocite{ofcomProtectingPeopleIllegal2023b}.

In terms of method, risks assessments for the purposes of the DSA are expected to look into --- amongst others --- the data collection and use practices, content moderation strategies, the online interface design, and risks of coordinated manipulation of their services (recitals 84--90). All of these areas increasingly include algorithmic systems in their functioning \autocites{gorwaAlgorithmicContentModeration2020}{leerssenSeeingWhatOthers2023}, particularly where minority languages are in use and firms lack staff specialising in those communities \autocites{authoriteitpersoonsgegevensAIAlgorithmicRisks2024}{koeneckeRacialDisparitiesAutomated2020}{okoloMakingAIExplainable2022}{pangAuditingCrossCulturalConsistency2023}. While these regulations require algorithmic assessments, they do not specify methodologies or approaches for understanding challenging, iterative systems, leaving a wide degree of flexibility to organisations \autocite{goodmanALGORITHMICAUDITINGCHASING2023}. Some organisations within scope of the DSA and OSA may have to deal with potentially illegal AI systems or models being uploaded to platforms, as a form of content, heightening the need for expertise in understanding them as the nature of user-uploaded content changes \autocite{gorwaModeratingModelMarketplaces2024}.

\subsection{Novel audit-like actors}\label{novel-audit-like-actors}

The DSA introduces novel audit-like mechanisms for large online platforms and entrust particular categories of entities and professionals with audit-like competencies and powers. In particular, two such categories under the DSA are the `vetted researchers' and the `trusted flaggers' --- both forms of `outsider oversight'. The former will be independent researchers affiliated with a research organisation and trusted to conduct specific research projects they have applied for in detect, identify, and understand the systemic risk posed by VLOPs'/VLOSEs' systems and services (art. 40 DSA). In particular, according to art. 40 of the DSA, the Digital Services Coordinator or the Commission will be responsible for initiating a data access request that will allow them and/or vetted researchers to monitor and assess compliance of the VLOP or VLOSE in question with the DSA \autocite{ausloosResearcherAccessDigitalforthcoming}. The OSA does not directly introduce a parallel category to vetted researchers --- an issue that was criticised in the Bill's passage \autocite{OnlineSafetyBill2023} --- however it obliges the regulator Ofcom to produce a report on researcher data access, and guidance for regulated services, within 18 months from the period when the Secretary of State commences that provision (s. 162). In the DSA, vetted researchers can request data, and platforms either have to provide it, or provide effective alternatives for study. The way that such access might extend to infrastructure --- important given the dynamic and difficult to locate `algorithms' within such platforms \autocite{seaverAlgorithmsCultureTactics2017} --- is, however, less clear \autocite{dariusImplementingDataAccess2023,dergachevaImprovingDataAccess2023,vealeDeniedDesignDataforthcoming}.

`Trusted flaggers' might be conceived of as a form of auditor --- organisations of particular expertise and competencies whose notices of illegal content will be dealt as matter of priority and processed by the online platform without undue delay. Their role is envisaged of one of both monitoring and of action, and again, it is the Digital Services Coordinator which can award this status, following the process set out in DSA art. 22. Such flaggers however will struggle to have impact at content moderation scale unless they opt for automation themselves, and in the DSA, do not a direct ability to scrutinise or influence algorithmic systems \autocite{appelmanTrustedFlaggers2022}.

In general, the audit-like powers of vetted researchers and trusted flaggers will be task-specific. In particular, the former's powers and scope of action will depend on the Digital Services Coordinator's reasoned request and the data requested therein, whereas trusted flaggers' audit-like powers will depend on the particular expertise of the accredited organisations. In practice, however, it is entirely unclear how challenges for the technical implementation of these powers will be overcome \autocite{dariusImplementingDataAccess2023,leerssenScrapingEuropeLaw2023}.

\subsection{External auditors}\label{external-auditors}

Finally, the DSA and the OSA introduce important provisions and mandates for the external auditing of large online platforms that, as discussed further in Sections \ref{potential-trajectories} and \ref{elements-of-the-emerging-political-economy-of-algorithmic-auditing}, set the foundations for the emergence of the industry of algorithmic auditing. In this direction, different from art. 40's externally led procedure for monitoring via researcher data access, art. 37 of the DSA sets out a form of annual, mandatory external auditing, to be performed by an independent auditing organisation which will be trusted with assessing with `a reasonable level of assurance' compliance with all DSA obligations (see discussion below in Section \ref{societal-implications-and-the-required-expertise}). The relevant provisions were recently supplemented by the DRPA which includes additional rules and operational details with regard to the performance of these audits.

In a similar but more irregular manner, the OSA introduces the `skilled person's report', commissioned research either by OFCOM or the provider under scrutiny at OFCOM's request and approval. The scope of such a report will be to help OFCOM in investigating a provider's (possible) failure to comply with OSA mandates and/or to understand the various risks related to the functioning of the provider's systems and services. Alongside this ability to order and commission a `skilled person's report', OSA empowers OFCOM to conduct its own inspections and audits as part, for example, of an investigation by appointing authorised persons and following the issuance of a respective notice to the provider (OSA s. 107 \& sch. 12). A flashpoint in the pipeline for skilled person's reports is any potential attempt to require client-side scanning in messaging or other services in relation to child-abuse imagery, a particularly controversial part of the OSA \autocite{lomasSecurityResearchersLatest2023}, but one which requires a skilled person's report before any such order can be made (OSA s. 122 (1) and OFCOM Consultation para. 28.22) \autocite{abelsonBugsOurPockets2021a}{AppleJoinsOpposition2023}{OpenLetterSecurity}. Insofar as such a report seeks to cove algorithmic detection technologies themselves, it may prove highly political, as previous UK Government funded reports from a research consortium scrutinising entrants to a `Safety Tech Challenge Fund' on child abuse detection in encrypted systems UK stimulated `safety tech' for encrypted environments found such tools unable to be effectively audited \autocite{peersmanFrameworkEvaluatingCSAM2023}.

\section{Potential Trajectories}\label{potential-trajectories}

Before delving into the foundations of the emerging political economy of algorithmic auditing, we wish to set out --- at a general level of abstraction --- three potential trajectories and scenarios for the future of algorithmic auditing. This intellectual exercise is admittedly arbitrary and perhaps oversimplified but in doing so, we wish to illustrate possible futures for transnational algorithmic governance in order to better explore the dynamics of the undergoing material, institutional, and socio-economic transformations \autocites{beckGovernanceSociotechnicalTransformations2021}{jasanoffDreamscapesModernitySociotechnical2015}. This is a loose futures exercise \autocite{hmgovernmentFuturesToolkitTools2017} which we deploy here in a narrative fashion to showcase and interrogate potential flavours of algorithmic auditing, as socio-technical praxis embedded within a system of legal rules, cultural norms, and material practices, and situated within the broader ecosystem of technology production.

\subsection{Convergence with traditional auditing}\label{convergence-with-traditional-auditing}

In this scenario, cultures and methodologies of traditional auditors dominate the market for, and practice of algorithmic auditing. Tech companies and AI start-ups in the field of model validation and assessment are absorbed by the Big 4 --- the world's major audit firms, Deloitte, EY, KPMG, and PwC --- thereby supplementing their auditing portfolios and augmenting the capabilities of their products and platforms. Nearly all large businesses in the US and UK use these firms: in 2017, 497 of the S\&P 500 in the United States, with PwC alone claiming to provide services to 87\% of the Fortune 500 \autocite[2]{gowBigFourCurious2018}. These firms have long sought to use legislation to craft lucrative roles for themselves --- both James Deloitte and Edwin Waterhouse (the `W' in PwC) coauthored Britain's Regulation of Railways Act 1868, which mandated a specified form of double-entry accounting that their firms specialised in --- unsurprisingly ending up as major railway auditors as a result \autocite[37]{gowBigFourCurious2018}. However, while these firms have been behind countless foundational statutes obliging audit to occur, they (and their industries bodies) have for over a century fiercely resisted any effort to legislate \textit{how} they should audit, with their judgment (and they way they choose to exercise it) seen as paramount in their eyes \autocite[57-8]{gowBigFourCurious2018}.

Thus, in this trajectory, the substantive elements of the auditing process, the benchmarks, and methodologies are determined by the negotiations of the Big 4 with their `clients' and little to no room is left for civil society organisations and researchers to meaningfully engage with, and be informed about the functioning of these systems \autocites{lauxTamingFewPlatform2021}{phiriSocialNetworksCorruption2019}. Eventually, like traditional audits, algorithmic auditing becomes just another service in the portfolios of multinational corporations conducting statutory audits --- just part of the furniture of modern capitalism  \autocite[55]{gowBigFourCurious2018}.

\subsection{Hybrid practices and hybrid, but separate, cultures}\label{hybrid-practices-and-hybrid-but-separate-cultures}

In this scenario, a vibrant start-up ecosystem for the assessment of algorithmic systems emerges. Leveraging its organisational agility and human capital, it overtakes the Big 4 in the offering of services for algorithmic auditing and offers a canvas of different methodologies for audited organisations to choose from. However, the autonomous development of these companies does not by itself suffice to safeguard their organisational independence. Instead, the erratic dynamics of AI-driven venture capital and the constant need for `healthy' balance sheets have a substantial impact on the direction of these companies. Financial incentives and strategies forge corporate relationships and routines of corporate practice that are both order-enabling, in that they create and stabilise legally constituted forms of private ordering with real-world effects as to how the `algorithmic society' is measured and evaluated, and norm-shaping, in that auditors and their sub-contractors produce, through their interaction, the substantive indicators based on which algorithmic systems will be eventually tested and measured.

\subsection{New forms, cultures, and actors of/for auditing}\label{new-forms-cultures-and-actors-offor-auditing}

In this scenario, research-driven, peer-reviewed, and evidence-based auditing methodologies rise as spin-offs from open-source communities, academic institutions, and non-profit organisations to become the new standards for the holistic evaluation of algorithmic systems \autocite{markdangelo2023}. Their organisational structure and operation are financially independent from the contractual relationship with the organisations they audit and the fees associated with the undertaken audit are re-directed to other non-profit organisation in the field. The academic credentials they carry along with the scholarly reputation of the people involved renders a collaboration with them strategically imperative for any company wishing to demonstrate its active commitment in building `responsible AI'. Academic research and journalism investigations are financially supported to become the main drivers of the field whilst civil society organisations, policy fora, and research communities provide the intellectual fuel, human capital, and technical expertise necessary for stabilising relevant routines of socio-technical practice.

\section{Elements of the Emerging Political Economy of Algorithmic Auditing}\label{elements-of-the-emerging-political-economy-of-algorithmic-auditing}

Where do we seem likely to end up amidst these scenarios? Within a universe of potential trajectories for social action and coordination, law is oftentimes imagined and understood as a system of rules for guiding human behaviour based on certain normative and procedural criteria such as human rights and the rule of law. Although engaging in a legal-theoretical debate remains well out of the scope of this paper, we believe it is important to highlight that our legal enquiry departs from the understanding of law as the form-giving institution that stabilises expectations and exchanges by weaving together the economy with the political system in its state form through formal, often --- and importantly for our case --- profession--based, organisational arrangements that produce collectively binding decisions within their respective functional areas \autocite[19-22]{kjaerLawPoliticalEconomy2020}. Oftentimes, and increasingly in the field of technology, in doing so it offloads normative work and routines of such practice to governance regimes that feed on the coordinated interactions of their subjects and imperatives of economic efficiency to (re)produce order that seems and operates beyond law \autocite{hartmannEvolutionIntermediaryInstitutions2015,wolframWritingRulesEurope2014}. Throughout this process, law-making as a practice of form-giving and form-shaping renders certain scenarios thinkable and others unthinkable; certain futures imaginable and others unimaginable \autocite{sobelreadReimaginingUnimaginableLaw2020}.

In the case of algorithmic auditing, law delegates much of its accountability monitoring to external auditors. Given the ambiguous effectiveness of internal audits and with vetted researchers and trusted flaggers assuming a delegated and task-specific auditing role, auditing work of the kind that civil society and the public at large need in order to understand and scrutinise algorithmic systems and services, is left at the discretion and powers of external auditors. Despite their central role in modern algorithmic governance, however, we argue that these actors will operate in a legal landscape whose institutional framework and safeguards are not robust enough to confront the powerful economic incentives and inter-organisational dependencies that can develop on the ground. Instead, a closer look at relevant provisions and industry developments may suffice to lower our expectations on the transformative potential of these new laws. In this direction, the provisions around the independence of auditing organisations, the selected standards and methodologies, the broader socio-economic questions that auditing is expected to encompass, as well as the operational details of the actual auditing, altogether create an institutional landscape which remains vulnerable to the stabilising effect of corporate strategies and routines of organisational practice. We will now examine these provisions in order.

\subsection{Independence is complex}\label{independence-is-complex}

One of the most crucial safeguards for the effective performance of an audit is the independence of those trusted to conduct it \autocite{costanzachockWhoAuditsAuditors2022}. The legal regimes we study here create confusion and uncertainty around this important audit characteristic.

\subsubsection{Unclear OSA Independence}
 Issues of independence are not deeply considered in the OSA. Much rests on OFCOM's nomination of the `skilled person' (or approval of the provider's recommendation for that matter) as to safeguard an independent investigation (OSA s 104 (4)-(6)). Furthermore, as both the issuance of a skilled person's report and the `powers of entry, inspection and audit' will be typically reserved for `more serious cases' (art. 28.52 OFCOM Consultation), questions of independence of the auditors entangle with the regulatory discretion of which cases warrant such interventions.

\subsubsection{DSA Audit Independence, But Blurred Compliance Roles}
 In contrast, the DSA establishes specific criteria of independence that VLOPs/VLOSEs are required to take into account when appointing an auditor (DSA art. 37 (3)). However, the audit process gets more complicated when considering the compliance function in art. 41, where an officer under that provision is responsible for both identifying and mitigating risks, but also organising and supervising any audits. The independence of the compliance officer (who is expected to be a senior manager employed by the provider or a contracted third party) is expected to be safeguarded by organisational and operational measures, whereas the independence of the auditing organisation by examining potential conflicts of interest (i.e.~by the provision of non-audit services related to the audited matters, such as software services, consultancy, training services, or content moderation services {[}recital 8, DPRA{]}) and ensuring the absence of fees `which are contingent on the result of the audit'(art. 37 (3) DSA). Yet we can see indications of which actors desire to claim their role in this space from the final sentence of recital 8 of the DRPA --- not part of the initial draft but which found its way to the final text --- which reads: `{[}Provisions on conflicts of interest{]} should not exclude auditing organisation who have performed statutory financial audits {[}for the audited organisation{]}'. This is a clear invitation to the so-called `Big 4' which have already declared their active interest and manifested their intentions to enter the industry of algorithmic audits \autocite{DeloitteIntroducesTrustworthy,EYTrustedAI,golbinAlgorithmicImpactAssessments,ResponsibleAIKPMG2023}. It is not difficult to imagine how the requirements of/for independence will play out in case a VLOP/VLOSE chooses to audit its financial and DSA obligations with the same auditor --- will PwC or EY think carefully before publishing a `negative' audit report for Meta's or Google's recommender systems?

\subsubsection{Independence Across the Supply Chain}
Auditors under the DSA can contract out part of the auditing process when it is necessary to seek expertise to evaluate, for example, `the design and functioning of algorithmic systems', `the risks to fundamental rights', or `the spread of illegal content' (recital 3 DRPA). The audited organisation remains responsible for ensuring the independence and expertise criteria are met (art. 4 DPRA), but it is unclear on what basis they will have information to do this, given information on the subcontractor and the reasons for their selection will be mediated through the initial auditor, and any subcontractors seem likely feel incentivised to remain within the contractual radar of this auditing ecosystem to ensure steady flow of future projects and partnerships, which may pressure them to give more positive comments than not. The extent to which an independent subcontractor could give a negative opinion which will be faithfully integrated, rather than buried, seems questionable. 

These issues echo similar concerns found in the audit of multinational groups, with regulators regularly raising issues about the performance and competence of so-called `component auditors' which a lead auditor may contract to perform a certain part of an audit, such as a particular national component of a corporate grouping \autocite{zhuHowLeadAuditor2024}. Literature on issues in component auditors already highlights the difficulties group auditors face in knowing and supervising components, even in cases where they are auditing something in a way that should be familiar to the group auditor (i.e., only the jurisdiction and tax systems differ) \autocite{sunderlandMultinationalGroupAudits2017}. It further highlights that group auditors already struggle with the sociocultural, regulatory, and institutional differences between them and their component auditors, even where these components are a separate entity using their own corporate umbrella and brandname (e.g. \textit{Deloitte Deutschland} or similar) \autocite{downeyChallengingGlobalGroup2021}. This seems to be even more difficult when the group auditor is subcontracting out a qualitatively different form of audit --- for example, of an algorithmic system, which they may not have expertise to interrogate or validate. 

Furthermore, algorithmic systems in areas like content moderation are increasingly servitised into networked supply chains and value chains with many actors and `many hands' \autocite{cobbeArtificialIntelligenceService2021,cobbeUnderstandingAccountabilityAlgorithmic2023}. Assessment of the auditors' and subcontractors' independence given these components --- such as content moderation technology providers such as Thorn, Google's Perspective API, Sightengine, Two Hat, and many more --- is not clearly guaranteed in the legislation and becomes extremely difficult to assess. In a world of AI supply chains, the reputation of the audited entities' components is perhaps even more important to consider as a corrupting factor than from the audited entity itself \autocite{lauxTamingFewPlatform2021}. The potential to subcontract draws a thread between the three scenarios illustrated in Section \ref{potential-trajectories}, allowing large auditors to retain functional control while using specific expertise of smaller organisations in the process, with unclear consequences for rigour or independence.

\subsection{Whose benchmarks? Whose methods?}\label{whose-benchmarks-whose-methods}

The baseline criteria and benchmarks based on which an algorithmic audit is (expected to be) performed are an integral part of an auditing process, and determinative of the nature and scope of the methodology \autocites[24]{bandyProblematicMachineBehavior2021a}[4]{brownAlgorithmAuditScoring2021}{rajiSavingFaceInvestigating2020}.
It can alter the way we think about the very thing we are expected to audit and measure \autocites{chandioHowAuditingMethodologies2023}{poonScorecardsDevicesConsumer2007}{poonNewDealInstitutions2009}.
It will not be news to most readers that translating high-level and generalised principles of fairness, bias, or transparency into an actionable and, by extension, auditable obligation can prove exceptionally challenging \autocites{goodmanAIAuditWashingAccountability2022}{blodgettLanguageTechnologyPower2020}{brownAlgorithmAuditScoring2021}{smithManyFacesFairness2023}.
Similarly, negotiating benchmarks for quality assessment internally can itself prove a laborious and demanding process as siloed teams can pose significant communication problems and inefficiencies \autocites{naharCollaborationChallengesBuilding2021}{holsteinImprovingFairnessMachine2019}{maffeyMLTEingModelsNegotiating2023}{sculleyHiddenTechnicalDebt2015}.
Unsurprisingly, there has been a plethora of approaches and toolkits for assessing algorithmic systems, with varied yardsticks for measurement and evaluation depending on the nature of the algorithmic system in question \autocites{balaynFairnessToolkitsCheckbox2023}{koeneckeRacialDisparitiesAutomated2020}{messmerAuditingRecommenderSystems2023}{norvalNavigatingAuditLandscape2023}{ulloaScalingSearchEngine2022}{wongSeeingToolkitHow2023}{damourFairnessNotStatic2020}.

Benchmarks can vary widely in terms of method, approach, and subject matter \autocites{rajiItTimeDevelop2022}{rosenbaumAlgorithmicAccountabilityDigital2019}. More holistic evaluations will likely draw on qualitative work, including interviews, database and document inspection. Instead, if it is simply the technical specifications of an AI model that are audited, then accepted quantitative methods that, for instance, produce a score for `robustness' may prove sufficient \autocite{chowdhuryWhatAudit}. Some audits rely on live interaction with the infrastructure, such as sock-puppet audits, i.e.~where a `fake' user is created and researchers observe the user's interaction with the platform \autocite{bartleyAuditingAlgorithmicBias2021}, or `bottom up', user-driven auditing \autocite{shenEverydayAlgorithmAuditing2021}, supported by researchers \autocite{bandyProblematicMachineBehavior2021a}. Others are more observational, taking slices of data or code, either from the organisation or via users, for in vitro analysis \autocite{ausloosResearchingDataRights2020,bandyProblematicMachineBehavior2021a}. These methods incur different costs (and reputational risks), but there are few, if any, studies directly comparing their efficacy and coverage in the context of social media. Effective methods may require creativity and novel methodological generation, rather than turning to an `accepted' benchmark or approach, particularly in the context of rapidly changing business models and platform practices. As a result, the choice of method is a commercial decision with unclear impacts on audit results.

Benchmarks and methods further reproduce the particular visions and worldview of the auditing entity. For example, METR (previously ARC Evals), a non-profit spin-off of the Alignment Research Center, evaluate AI systems based on their capabilities for `autonomous replication', meaning whether `an AI could survive on a cloud server, obtain money and compute resources, and use those resources to make more copies of itself' \autocite{ARCEvalsMETR}. In a similar spirit, Apollo Research (or AI Evals), a project sponsored by the Rethink Priorities Initiative, aims at building evaluation models to detect `deception and potentially other misaligned behavior' \autocite{ApolloResearch,hobbhahnAnnouncingApolloResearch}. In contrast, and indicatively, Fiddler Auditor tests systems for model robustness based on principles such as transparency, interpretability, fairness, privacy and reliability \autocite{FiddlerAIAI}. This becomes important, as the breadth of systemic risks that the DSA requires to be considered also allows considerable room for framing, bringing the auditor's priorities and worldview and constructing the DSA requirements around them, rather than the other way around. Pushing back against this may be hard for an under resourced Digital Services Coordinator, similarly overwhelmed by the potential breadth of risks that, in slickly written text, seem genuine, that could be considered as part of such an audit.

Under the DRPA, responsibility for the initial formulation of the benchmarks against which compliance is/will be sought, rests with the audited organisation (DRPA art. 5 (1) (a)) and in practice, it is more likely than not that this process will be carried out by the compliance function that presented above (DSA art. 41 and recital 99). Importantly, the evaluation of these benchmarks by the auditing provider is unlikely to have a negative impact on the final outcome of the auditing. Instead, as recital 16 and art. 8 of the DRPA explicitly acknowledge, the audit conclusion should be `positive with comments' when the auditing organisation considers it necessary to provide further comments on the selected benchmarks in order to `usefully inform' the future `benchmarking' of the audited organisation, based on the auditor's knowledge, research, and expertise. This `agile' back-and-forth between the auditing organisation and the (compliance function of the) audited organisation is likely to have a considerable impact on the way auditing benchmarks are formulated and/or standardised in the future. As the inter-organisational benchmarks consolidate overtime in a form of `benchmark-as-we-go', the relationship between the `audited' and the `auditing' may generate undesirable dependencies \autocite{rakovaWhereResponsibleAI2021}. In this interplay, the primary responsibility for the formulation of benchmarks is left to the auditing organisation and the only avenue available for external input in the process is the latter's discretion of using `information from external sources' in its `positive with comments' audit opinion (recital 16 DRPA). Eventually, benchmark disparities amongst different auditors may incentivise platforms to choose their assessors and auditors based on their benchmarks (easy or difficult, simple or complicated) and/or intensify institutional and organisational dynamics towards benchmark standardisation, a process that will be in itself extremely difficult to navigate and deliver \autocite{nanniniExplainabilityAIPolicies2023}.

\subsection{Societal implications and the required expertise}\label{societal-implications-and-the-required-expertise}

Several studies have discussed the necessity yet complexity of incorporating societal considerations in the assessment of algorithmic systems \autocites{balaynFairnessToolkitsCheckbox2023}{balaynDebiasingRegulatingAI2021}{mitchellAlgorithmicFairnessChoices2021}. The processes for doing so and the effect that organisational structures and practises can have on them are equally well researched and documented empirically \autocite{madaioCoDesigningChecklistsUnderstand2020}. The `systemic risk' analysis VLOPS/VLOSEs are subject to in the DSA requires such analysis, an obligation fortified by the DRPA, obliging external audits to take into account --- amongst other factors --- the nature of the audited service and `the societal and economic context in which the audited service is operated' (DRPA art. 9(4)(a)). When considering this context, the auditing organisation is required to express an opinion with a `reasonable level of assurance' (DRPA art. 3 and recital 16). Reasonable assurance has no firm definition, but sits in contrast and is weaker than `absolute' assurance, indicating the auditor is not a guarantor of correctness, and even audits conducted in accordance with given standards may fail to detect material concerns \autocite{elderAuditingAssuranceServices2020}. In financial audits, this stems from sampling, complex estimates that can be changed by fast-moving events, the potential for sophisticated fraud, and the need to make audits economically viable \autocite{elderAuditingAssuranceServices2020}.

Unsurprisingly, large multinationals' consultation contributions concerning the DPRA indicates that this level of assurance, which survived to the final version of the DPRA, is one they are unhappy with. Google argued that the novelty of the field and lack of well-stablished rules leads to a material risk of `the reasonable level of assurance standard' being interpreted and applied inconsistently by different auditing actors \autocite{googleSubmissionsGoogleResponse2023}. In a similar context, Booking.com argues the reasonable assurance standard in other sectors `is generally reserved for subject matters that are highly quantitative (and even binary) in nature' \autocite{booking.comSubmissionsBookingCom2023}.

There are clear winners in all of this however --- the Big 4, again. Part of the reason why a `reasonable level of assurance' is hard to explain is because over the last century, `large accounting firms used their links with regulators and standard-setters to co-produce a coded, excluding but otherwise benign professional language that is rich with acronyms, jargon and euphemisms' \autocite[59]{gowBigFourCurious2018}. Gow and Kells argue terms like `reasonable assurance' deliberately `border on nonsense' \autocite[59]{gowBigFourCurious2018}. While auditing has become more formalised, with rules describing their form and structure, the discretion has simply been condensed into these arcane and opaque terms, with standards laying out the banal --- that audits should have a title page; that they should list the client's instructions, and so on. Consider (as Gow and Kells do) the following description of a `limited assurance engagement' in a major Australian audit standard:

\begin{quote} {\itshape
An assurance engagement in which the assurance practitioner reduces engagement risk to a level that is acceptable in the circumstances of the engagement, but where that risk is greater than for a reasonable assurance engagement, as the basis for expressing a conclusion in a form that conveys whether, based on the procedures performed and evidence obtained, a matter(s) has come to the assurance practitioner’s attention to cause the assurance practitioner to believe the subject matter information or subject matter is materially misstated. The nature, timing and extent of procedures performed in a limited assurance engagement is limited compared with that necessary in a reasonable assurance engagement but is planned to obtain a level of assurance that is, in the assurance practitioner’s professional judgement, meaningful. To be meaningful, the level of assurance obtained by the assurance practitioner is likely to enhance the intended users’ confidence about the subject matter information or subject matter to a degree that is clearly more than inconsequential. } \autocite[59-60]{gowBigFourCurious2018}
\end{quote}

If you think that makes little sense --- we agree. 

Algorithmic auditing takes the now well-known problem of having standards set in an excruciatingly arcane and opaque language to another level, creating a significant new tension in the process. Auditing organisations are expected to assess contextual, societal features, which requires creative and ambitious methods. Yet they should root their analysis in `proven expertise in the area of risk management, technical competence, and capabilities' (art. 37(3)(b) DSA). Traditional audit methodologies are standardised, repeatable, even if the ways in which they are standardised and repeated are closed industry knowledge. The types of audits that seem to be required by the DSA texts, if they are to be rigorous, are necessarily not.

This does indeed make the standard of the `reasonable level of assurance' difficult --- not because of its ambition of rigour, but because this definition is rooted in the idea of an institutional field, shared (yet often proprietary) understandings, methods and norms between professionals and their organisations in this space \autocite{bakerRegulationStatutoryAuditing2014,dimaggioIronCageRevisited1983}. The necessity of contextual, creative analysis leaves key questions to the professional discretion --- and socio-political vision --- of the auditor \autocites{goodwinProfessionalVision1994}{metcalfOwningEthicsCorporate}. This may not inherently be a bad thing, as long as assumptions are placed on the table. Yet the DSA, and the internal compliance roles it envisages, seem to have few incentives to push auditors away from rote, unambitious, context-free standards towards the creativity and creative rigour needed to undertake sociotechnical analysis with an open-ended list of potential systemic risks. The history of the audit industry indicates that auditors will fiercely protect their processes from external influence. In this context, auditing benchmark and methods seem likely to become just another standard(ised) service in the portfolio of traditional auditing actors from the Global North with their often monolithic cultures and one-dimensional methodologies of/for auditing.

\subsubsection{A concerning statutory pressure valve}
More worryingly, we can see legally that in the face of an audit system that desires standardisation and repeatability in order to provide `assurance', auditors are likely to lean on an exit clause in the DSA which enables them to escape complex, value-laden, socio-technical analysis. art. 37(5) of the DSA enables the auditing organisation to avoid auditing specific elements or expressing an opinion on certain aspects of an investigation as long as it includes `an explanation of the circumstances and the reasons why those elements could not be audited'. Rather than try to reinvent the concept of reasonable assurance such that open-ended challenges can face external scrutiny, as a superficial reading of the DSA might seem to push for, the existence of art. 37(5) provides a way to claim that such areas are simply too hard to analyse, giving a false level of assurance by letting the auditing organisation effectively scope out the qualitative, and potentially most societally crucial, aspects of their mandate.

\subsection{Early signs of a new (or old) industry}\label{early-signs-of-a-new-or-old-industry}

Today, as the DSA enters into force, industry dynamics have already started taking shape. PwC has launched its Responsible AI Toolkit aiming at offering an `end-to-end enterprise governance framework' to enable oversight and traceability of a company's AI development lifecycle. EY pitches its `Trusted AI' tool as a platform capable of providing insights and helping AI design teams in `quantifying risks' \autocite{EYTrustedAI}. KPMG's `Responsible AI' is advertised as a form of AI governance that exposes risks and vulnerabilities without `compromising on innovation' \autocite{ResponsibleAIKPMG2023}. Deloitte's `Trustworthy AI Framework' promises to infuse `an ethical mindset within {[}an{]} organization' by --- amongst others --- engaging `external ethics experts and academic institutions to conduct well-rounded client conversations' \autocite{DeloitteIntroducesTrustworthy}. Meanwhile, industry partnerships are mushrooming. KPMG and Microsoft have signed a multibillion-dollar agreement for the expansion of their relationship that `will reshape professional services {[}\ldots{]} including {[}\ldots{]} {[}the{]} use of Artificial Intelligence solutions for clients, industries and society more broadly'. Deloitte has teamed up with ChatterboxLabs, an AI startup, to enhance model insights offered through its `Trusted AI toolkit' \autocite{DeloitteAIInstitute}. Arc Evals has partnered with Anthropic and OpenAI to evaluate their AI systems \autocite{ARCEvalsMETR}.

In parallel, AI auditing start-ups are attracting finance in the form of donations and venture capital; early signs of a burgeoning ecosystem \autocite{goodmanALGORITHMICAUDITINGCHASING2023,kayeAIRegulationCoulda}. These are usually companies specialising in testing the reliability and resilience of an AI model, performing root cause analysis in Large Language Models (LLMs) and troubleshooting for model drifts, and generally monitoring and observing the model's performance. Examples start-ups that have received seed funding include Truera, Arize, Arthur, Hollistic AI, Babl AI, and Aporia whereas Apollo Research and METR, as we have already seen, have spun off from lucrative research groups and initiatives as non-profits. It is highly likely that most of the aforementioned start-ups as well as the traditional actors in the field will strive to claim role not only in the DSA Audit space but potentially in simultaneously in AI conformity assessments under the AI Act (even though third party audits are very, very rarely required by the proposed Regulation) \autocite{bablaiAIAudits,DigitalServicesAct,EUAIAct}. These early developments seem to indicate a process of professionalisation not dissimilar to the industry transformations that have taken place in the fields of corporate sustainability, finance and accounting \autocites{botzemPoliticsAccountingRegulation2012}{humphreyGlobalAuditProfession2009}{matusCertificationSystemsMachine2022}. However, given the sheer size of the incumbent auditing players, and their history of aristocratic-style marriages and mergers to consolidate into the concentrated Big 4 we see today \autocite{gowBigFourCurious2018}, we do have to ask how long such small players will even remain independent entities.

\subsection{A sharp departure from idealised research-led audits}\label{a-sharp-departure-from-idealised-research-led-audits}

The terminology `audit' has a non-conventional use in the academic research community studying algorithmic systems. It has often been used to indicate a study or investigation carried out by researchers, non-profits or by journalists \autocite{birhaneAuditingSaliencyCropping2022,sandvigAuditingAlgorithmsResearch2014}. It has been understood as `similar in spirit to the well-established practice of bug bounties', even in works that clearly contextualise audits in the context of their practices in other industries \autocite{rajiClosingAIAccountability2020}. The observations from the emerging political economy of algorithmic auditing above pose challenges for this understanding and this paradigm. This requires particular care to ensure that the term `audit', with the legitimacy awarded by scholarly communities such as FAccT, is not co-opted to mean something entirely different in practice \autocite{youngConfrontingPowerCorporate2022}. At the same time, it requires consideration of how, if at all, this flavour of audit might become sustainable --- more than an occasional research project using academics and journalists that soon, due to the incentives of their own roles, move to other challenges.

We have seen that in the world of audit, money and expertise moves erratically, and organisational dependencies, routines of corporate practices, and financial incentives can prevail. Such factors risk hindering the translation of the best analysis methods from scholarship to practice. The most challenging, yet most critical, socio-technical approaches may struggle to penetrate structures tailored to the intentions or desires of the audited or auditing entities \autocites{naharCollaborationChallengesBuilding2021}{holsteinImprovingFairnessMachine2019}. Financialisation and professionalisation may transform an inherently normative enquiry into an iteratively mundane practice of/for calculable, quantified deliverables. The regimes implied by the text of the DSA and OSA further push this direction. There might even be an element of complacency to this --- so far, discussions of idealised, methodologically rich and creative algorithmic audits have rarely been linked to potential legislative or regulatory requirements. It is true that such creative audits have been promoted, including as `soft' industry practices, but with less thought towards how they might practically be mandated.

Furthermore, a focus on public sector algorithmic systems in areas such as policing or welfare might have created confusion in this space. Where audits apply to discrete projects in public bodies, these might be one-off reports where the efficacy and appropriateness can be assessed by a court in case of a judicial review or similar public law challenge. For example, the audit documents accompanying live facial recognition tools in response to the \textit{Bridges} case in England and Wales \autocite{EqualityImpactAssessment,PolicyDocumentOvert,BridgesChiefConstable2020,StandardOperatingProcedure2022,UnderstandingAccuracyBias2022} or accompanying the England and Wales Ofqual COVID-19 exam results algorithm \autocite{AwardingGCSELevel}, did not follow a set standard, instead resembling more ad-hoc quality assurance and modelling criteria familiar from decades of government analytics and modelling practice \autocite{hmtreasuryAquaBookGuidance2015,pbluncertainty,vealeAdministrationAlgorithmPublic2019}. But companies, compliance departments, audit industries, and insurers, do not operate in the same way as a public body can when using discretion when discharging their functions If audits are to be carried out regularly, across an entire industry, they institutionally crystallise in ways that make them distinct from the uses of societal or ethical analysis in high-stakes public sector machine learning contexts. This is not to say that public bodies do not find roles professionalised or institutionalised, but that rationality or reasonableness tests lend themselves can, in some ways, lend themselves to more flexible approaches, compared to logics of large multi-national business structures.

\section{Foundations for Better Audit Today}\label{foundations-for-better-audit-today}

Despite the above, we want to end by highlighting several directions already in the DSA and the OSA which might, if given sharp attention by Digital Services Coordinators, Ofcom or the Commission, help mitigate some of the issues and challenges we have outlined above. Wholesale regulatory change is unlikely at this early stage, so our recommendations focus on what can be done immediately in these early, formative stages of a new regime.

\subsubsection{Transparency}
The first practicality refers to the transparency and identities of the people involved in the auditing process. A sentence that made it intact to the final text of the DRPA --- despite consultation arguments from companies including Google --- is art. 7(2). This Article provides that the agreement between the auditor and the audited organisation along with `any other agreements or engagements letters {[}between them{]}' that are relevant to the performance of the audit shall be annexed to the final audit report. In parallel, according to both art. 7 and recital 15 of the DRPA, the auditing organisation is required to specify the details of the staff responsible for carrying out the audit. Such a direction is also hinted in the `skilled person' profile envisaged by the OSA. Increasing the transparency around more than just the firm or the lead auditor, but the team, and entire supply chain, might go some way to help accountability issues, but only if regulators are willing to provide pressure and scrutiny on the disciplinary mix and methodological capabilities of the teams involved. This could be further supported by guidelines or informal pressure on the direction of reports from year-to-year. There is also the possibility of undertaking analysis on the tendencies or impartiality of these audit reports, to consider if certain auditors systematically underplay certain risks, and then to provide that as aggregate feedback. Such a focus highlights the role of `regulatory intermediaries' such as these auditors, both in general and in the algorithmic context \autocites{abbottTheorizingRegulatoryIntermediaries2017}{matusCertificationSystemsMachine2022}. Rules on transparency at the level of auditing organisations are already set out for financial auditors in EU law (Directive 2014/56/EU), yet these would not apply to organisations either solely carrying out, or in respect of, algorithmic audits. As mentioned above, subcontractors are a particular liability in this regard. Legislators may wish to consider such changes depending on the way the industry grows.

\subsubsection{Confidentiality}
Secondly, contrary to financial auditing that feeds on public statements made by the audited organisations, algorithmic auditing will inevitably deal with information kept in private. In particular, according to art. 5(2) of the DPRA, the audited organisation will provide a wealth of information to the auditor including but not limited to information on `decision-making structures, {[}\ldots{]} relevant IT systems, data sources, {[}\ldots{]} as well as explanations of relevant algorithmic systems and their interactions', as well as `all data necessary for the performance of the audit', including documents, testing environments as well as `personnel and premises of that provider, and any relevant sub-contractors'. There are explicit references to protections of confidentiality and trade secrets aimed at protecting commercially sensitive information of the audited organisation (art. 37(2) DSA and OFCOM Consultation 28.44-28.48), with Ofcom promising to remain `mindful of the importance of protecting {[}such information{]}' (OFCOM Consultation 28.48){]}.

Under art. 37(2) DSA, requirements of confidentiality `shall not adversely affect the performance of the audits'. Yet given how such information might reveal know-how both on underlying services and on the specific risk mitigation factors which other clients may benefit from their own auditors having knowledge of, it is easy to imagine organisations seeking to withhold commercially relevant information or seek extremely strong guarantees, where possible. Unlike in financial audits, where the base documents that may be required are likely to be more rote and predictable than not, the context--specific nature of sociotechnical algorithmic audits indicate that rigorous auditors need to both work out what documents might exist, and push for their release. Given the independence challenges discussed above, it is difficult to imagine this occurring with much ferocity. This is only compounded by art. 37(2) of the DSA, allowing auditors to publish a report while redacting what in their view is `reasonably be considered to be confidential'.

Regulators have the space to issue guidance on this area that auditors can --- and perhaps must --- lean upon to do their work. As they gain more knowledge of the types of risks and modalities of analysis, they should publish types of information they expect auditors to have unfettered access to, and require auditors to list documents or other resources they sought access to but were refused. Such small moves might tip balances inside audited--auditing organisation relations to create more externally accountable patterns of engagement, particularly as confidentiality, and especially trade secrets, can be asserted with very limited recourse to externally check whether these claims are even grounded in law.

\subsubsection{Timeframe}
Finally, in line with the DSA's annual lifecycle for risk assessment, audits are expected to cover a period of one year (DSA art. 34). Given the highly programmable nature of information and computational production, having an annual audit review instead of a `point-in-time' investigation clearly makes sense as it allows auditors to have a more holistic picture of a system's development lifecycle instead of merely capturing a snapshot of its functioning. However, precisely because of the amount of information that auditors will be expected to process and review, the agile way of internally negotiating and performing changes in an algorithmic system, as well as the tight timelines within which VLOPs/VLOSEs operate, auditors may confront an insurmountable amount of evidence to go through and not enough time. In this regard, neither the DSA nor the DPRA specify an indicative timeframe for the completion of the audit which is left to the discretion of the parties to determine (DRPA art. 7 (1)(d)). We believe that they should, or at least indicate a framework through which such a timeframe can be accountably determined.

\section{Conclusion}\label{conclusion}

Algorithmic auditing (and the closely associated conformity assessment standardisation `market' for AI systems) is here to stay --- but its form matters and it not yet set in stone. After almost a decade of academic and policy dialogue, we can now witness early signs of the real-world transformation. That is already a big step, but we must not get complacent. A systematic look on the relevant legislative provisions and the undergoing industry developments allows us to foresee (perhaps `with a reasonable level of assurance'?) the actors involved in the burgeoning field of algorithmic auditing as well as the routines and cultures of professional practice that are likely to shape the field's future. In an ecosystem driven largely by a blend of traditional auditors, standardisation dynamics, share prices and venture capital, law is inevitably translated into (inter-)organisational projects. In this direction, audit independence, benchmark and methodological selection, the evaluation of the socio-economic context of the algorithm, and the disclosure of data to the auditor, are all becoming increasingly stabilised objects of/for negotiations amongst interdependent organisations. What used to be a space for research and contestation, thus transforms into a set of deliverables for professionals. Witnessing this reality unfolding might give researchers in the field a good reason to reflect on, and review the way they study, talk, and (co-) think about `fairness, accountability, and transparency' in relation to the political economy of technology, and whose vision of `auditing' they have in mind when they use that term. There is hope --- and a legislative hook --- for better practices of audit, but they cannot be taken for granted. In this paper, we have sought to lay out some practical critiques and directions for the near-term development and potential repositioning of this field, and hope that audit proponents can take these insights to make strategically powerful interventions in their own jurisdictions and spheres of scholarly and applied influence.

\begin{acks}
This work is supported by the \grantsponsor{UKRI}{UKRI}{} under Grant
No.:~\grantnum[https://tas.ac.uk ]{EP/V00784X/1}{EP/V00784X/1} Trustworthy Autonomous Systems Hub.
\end{acks}

\printbibliography

@article{humphreyGlobalAuditProfession2009,
  title = {The Global Audit Profession and the International Financial Architecture: {{Understanding}} Regulatory Relationships at a Time of Financial Crisis},
  shorttitle = {The Global Audit Profession and the International Financial Architecture},
  author = {Humphrey, Christopher and Loft, Anne and Woods, Margaret},
  date = {2009},
  journaltitle = {Accounting, Organizations and Society},
  volume = {34},
  number = {6-7},
  pages = {810--825},
  doi = {10.1016/j.aos.2009.06.003}
}

@article{abbottTheorizingRegulatoryIntermediaries2017,
  title = {Theorizing {{Regulatory Intermediaries}}: {{The RIT Model}}},
  shorttitle = {Theorizing {{Regulatory Intermediaries}}},
  author = {Abbott, Kenneth W. and Levi-Faur, David and Snidal, Duncan},
  date = {2017-03-01},
  journaltitle = {The ANNALS of the American Academy of Political and Social Science},
  volume = {670},
  number = {1},
  pages = {14--35},
  doi = {10/f953b4}
}

@article{beckGovernanceSociotechnicalTransformations2021,
  title = {The Governance of Sociotechnical Transformations to Sustainability},
  author = {Beck, Silke and Jasanoff, Sheila and Stirling, Andy and Polzin, Christine},
  date = {2021-04},
  journaltitle = {Current Opinion in Environmental Sustainability},
  volume = {49},
  pages = {143--152},
  doi = {10.1016/j.cosust.2021.04.010}
}

@book{jasanoffDreamscapesModernitySociotechnical2015,
  title = {Dreamscapes of {{Modernity}}: {{Sociotechnical Imaginaries}} and the {{Fabrication}} of {{Power}}},
  shorttitle = {Dreamscapes of {{Modernity}}},
  author = {Jasanoff, Sheila and Kim, Sang-Hyun},
  date = {2015},
  publisher = {University of Chicago Press},
  location = {Chicago, IL},
  doi = {10.7208/chicago/9780226276663.001.0001}
}

@online{goodmanAIAuditWashingAccountability2022,
  title = {{{AI Audit-Washing}} and {{Accountability}}},
  author = {Goodman, Ellen P and Tr\'{e}hu, Julia},
  date = {2022-11},
  url = {https://www.gmfus.org/news/ai-audit-washing-and-accountability},
  urldate = {2024-01-11},
  organization = {German Marshall Fund of the United States}
}

@legislation{AlgorithmicAccountabilityAct2023,
  title = {Algorithmic {{Accountability Act}} of 2023},
  date = {2023-09-21},
  journaltitle = {U.S.C.},
  volume = {5},
  number = {H.R. 5628},
  pages = {553},
  publisher = {U.S. Government Publishing Office},
  url = {https://www.govinfo.gov/app/details/BILLS-118hr5628ih},
  urldate = {2024-01-18}
}

@online{ApolloResearch,
  title = {Apollo {{Research}}},
  url = {https://www.apolloresearch.ai/},
  urldate = {2024-01-16},
}

@article{appelmanTrustedFlaggers2022,
  title = {On "{{Trusted}}" {{Flaggers}}},
  shorttitle = {On "{{Trusted}}" {{Flaggers}}},
  author = {Appelman, Naomi and Leerssen, Paddy},
  date = {2022},
  journaltitle = {Yale J.L. \& Tech.},
  volume = {24},
  number = {1},
  pages = {452--475},
  url = {https://yjolt.org/trusted-flaggers}
}

@online{AppleJoinsOpposition2023,
  title = {Apple Joins Opposition to Encrypted Message App Scanning},
  date = {2023-06-27},
  url = {https://www.bbc.com/news/technology-66028773},
  urldate = {2024-01-16},
  organization = {BBC News}
}

@online{ARCEvalsMETR,
  title = {{{ARC Evals}} - {{METR}}},
  url = {https://metr.org/},
  urldate = {2024-01-01}
}

@book{ausloosResearcherAccessDigitalforthcoming,
  title = {Researcher {{Access}} to {{Digital Infrastructures}}},
  editor = {Ausloos, Jef and family=Souza, given=Siddharth, prefix=de, useprefix=true},
  year = {forthcoming},
  publisher = {Cambridge University Press},
  location = {Cambridge}
}

@article{ausloosResearchingDataRights2020,
  title = {Researching with {{Data Rights}}},
  author = {Ausloos, Jef and Veale, Michael},
  date = {2020},
  journaltitle = {Technology and Regulation},
  number = {1},
  pages = {136--157},
  urldate = {2021-01-04},
  doi = {10.26116/techreg.2020.010 }
}

@online{australiangovernmentRoyalCommissionRobodebt2023,
  title = {Royal {{Commission}} into the {{Robodebt Scheme}}},
  author = {{Australian Government}},
  date = {2023-11},
  url = {https://www.pmc.gov.au/sites/default/files/resource/download/gov-response-royal-commission-robodebt-scheme.pdf}
}

@online{authoriteitpersoonsgegevensAIAlgorithmicRisks2024,
  title = {{{AI}} \& {{Algorithmic Risks Report Netherlands}} - {{Winter}} 2023 2024},
  author = {{Authoriteit Persoonsgegevens}},
  date = {2024-01-18},
  url = {https://perma.cc/M2JU-GCVU},
  organization = {AP}
}

@report{AwardingGCSELevel,
  title = {Awarding {{GCSE}}, {{AS}}, {{A}} Level, Advanced Extension Awards and Extended Project Qualifications in Summer 2020: Interim Report},
  institution = {Ofqual/20/6656/1},
  author = {{Ofqual}},
  month = aug,
  year = 2020,
  url = {https://perma.cc/8QY4-VW4G}
}

@online{bablaiAIAudits,
  title = {{{AI Audits}}},
  author = {{BABL AI}},
  url = {https://babl.ai/ai-audits/},
  urldate = {2024-01-16}
}

@article{bakerRegulationStatutoryAuditing2014,
  title = {The Regulation of Statutory Auditing: An Institutional Theory Approach},
  shorttitle = {The Regulation of Statutory Auditing},
  author = {Baker, C. Richard and B\'{e}dard, Jean and Prat dit Hauret, Christian},
  date = {2014-01-01},
  journaltitle = {Managerial Auditing Journal},
  volume = {29},
  number = {5},
  pages = {371--394},
  publisher = {Emerald Group Publishing Limited},
  doi = {10.1108/MAJ-09-2013-0931}
}

@inproceedings{balaynFairnessToolkitsCheckbox2023,
  title = {“\checkmark{} {{Fairness Toolkits}}, {{A Checkbox Culture}}?” {{On}} the {{Factors}} That {{Fragment Developer Practices}} in {{Handling Algorithmic Harms}}},
  shorttitle = {“\checkmark{}  {{Fairness Toolkits}}, {{A Checkbox Culture}}?},
  booktitle = {Proceedings of the 2023 {{AAAI}}/{{ACM Conference}} on {{AI}}, {{Ethics}}, and {{Society}}},
  author = {Balayn, Agathe and Yurrita, Mireia and Yang, Jie and Gadiraju, Ujwal},
  date = {2023-08-29},
  series = {{{AIES}} '23},
  pages = {482--495},
  publisher = {Association for Computing Machinery},
  location = {New York, NY, USA},
  doi = {10.1145/3600211.3604674},
}

@article{bandyProblematicMachineBehavior2021a,
  title = {Problematic {{Machine Behavior}}: {{A Systematic Literature Review}} of {{Algorithm Audits}}},
  shorttitle = {Problematic {{Machine Behavior}}},
  author = {Bandy, Jack},
  date = {2021-04-22},
  journaltitle = {Proc. ACM Hum.-Comput. Interact.},
  volume = {5},
  pages = {74:1--74:34},
  doi = {10.1145/3449148},
  issue = {CSCW1}
}

@inproceedings{bartleyAuditingAlgorithmicBias2021,
  title = {{Auditing Algorithmic Bias on Twitter}},
  author = {Bartley, Nathan and Abeliuk, Andres and Ferrara, Emilio and Lerman, Kristina},
  date = {2021},
  pages = {65--73},
  eventtitle = {Proceedings of the 13th {{ACM Web Science Conference}} 2021}
}

@inproceedings{birhaneAuditingSaliencyCropping2022,
  title = {Auditing {{Saliency Cropping Algorithms}}},
  author = {Birhane, Abeba and Prabhu, Vinay Uday and Whaley, John},
  date = {2022},
  pages = {4051--4059},
  url = {https://openaccess.thecvf.com/content/WACV2022/html/Birhane_Auditing_Saliency_Cropping_Algorithms_WACV_2022_paper.html},
  urldate = {2024-01-15},
  eventtitle = {Proceedings of the {{IEEE}}/{{CVF Winter Conference}} on {{Applications}} of {{Computer Vision}}}
}

@inproceedings{blodgettLanguageTechnologyPower2020,
  title = {Language ({{Technology}}) Is {{Power}}: {{A Critical Survey}} of “{{Bias}}” in {{NLP}}},
  shorttitle = {Language ({{Technology}}) Is {{Power}}},
  booktitle = {Proceedings of the 58th {{Annual Meeting}} of the {{Association}} for {{Computational Linguistics}}},
  author = {Blodgett, Su Lin and Barocas, Solon and Daum\'{e} III, Hal and Wallach, Hanna},
  date = {2020},
  pages = {5454--5476},
  publisher = {Association for Computational Linguistics},
  location = {Online},
  doi = {10.18653/v1/2020.acl-main.485},
  eventtitle = {Proceedings of the 58th {{Annual Meeting}} of the {{Association}} for {{Computational Linguistics}}}
}

@report{booking.comSubmissionsBookingCom2023,
  title = {Submissions by {{Booking}}.Com in Response to the {{European Commission}}’s Request for Feedback on the Draft {{DRPA}}},
  author = {{Booking.com}},
  date = {2023-06-02},
  url = {https://ec.europa.eu/info/law/better-regulation/have-your-say/initiatives/13626-Digital-Services-Act-conducting-independent-audits/F3424053_en},
  urldate = {2023-12-09}
}

@online{BridgesChiefConstable2020,
  title = {R ({{Bridges}}) v. {{Chief Constable}} of {{South Wales Police}} {{[2020] EWCA Civ 1058}}},
  year = 2020,
  url = {https://caselaw.nationalarchives.gov.uk/ewca/civ/2020/1058}
}

@article{brownAlgorithmAuditScoring2021,
  title = {The Algorithm Audit: {{Scoring}} the Algorithms That Score Us},
  author = {Brown, Shea and Davidovic, Jovana and Hasan, Ali},
  date = {2021},
  journaltitle = {Big Data \& Society},
  volume = {8},
  number = {1},
  pages = {2053951720983865},
  publisher = {SAGE}
}

@book{brownRegulatingCodeGood2013,
  title = {Regulating {{Code}}: {{Good Governance}} and {{Better Regulation}} in the {{Information Age}}},
  shorttitle = {Regulating {{Code}}},
  author = {Brown, Ian and Marsden, Christopher T.},
  date = {2013},
  publisher = {MIT Press},
  location = {Cambridge, MA},
}

@article{marsdenPlatformValuesDemocratic2020a,
  title = {Platform Values and Democratic Elections: {{How}} Can the Law Regulate Digital Disinformation?},
  shorttitle = {Platform Values and Democratic Elections},
  author = {Marsden, Chris and Meyer, Trisha and Brown, Ian},
  date = {2020-04-01},
  journaltitle = {Computer Law \& Security Review},
  shortjournal = {Computer Law \& Security Review},
  volume = {36},
  pages = {105373},
  doi = {10.1016/j.clsr.2019.105373}
}

@online{ojewaleAIAccountabilityInfrastructure2024,
  title = {Towards {{AI Accountability Infrastructure}}: {{Gaps}} and {{Opportunities}} in {{AI Audit Tooling}}},
  shorttitle = {Towards {{AI Accountability Infrastructure}}},
  author = {Ojewale, Victor and Steed, Ryan and Vecchione, Briana and Birhane, Abeba and Raji, Inioluwa Deborah},
  date = {2024-03-14},
  eprint = {2402.17861},
  eprinttype = {arxiv},
  doi = {10.48550/arXiv.2402.17861},
}

@book{gowBigFourCurious2018,
  title = {The {{Big Four}}: {{The Curious Past}} and {{Perilous Future}} of the {{Global Accounting Monopoly}}},
  shorttitle = {The {{Big Four}}},
  author = {Gow, Ian D. and Kells, Stuart},
  date = {2018},
  publisher = {Berrett-Koehler Publishers, a BK Business book},
  location = {Oakland},
}

@article{buitenDigitalServicesAct2022,
  title = {The {{Digital Services Act}}: {{From Intermediary Liability}} to {{Platform Regulation}}},
  shorttitle = {The {{Digital Services Act}}},
  author = {Buiten, Miriam C.},
  date = {2022},
  journaltitle = {JIPITEC},
  volume = {12},
  number = {5},
  url = {https://www.jipitec.eu/issues/jipitec-12-5-2021/5491}
}

@article{chandioHowAuditingMethodologies2023,
  title = {How {{Auditing Methodologies Can Impact Our Understanding}} of {{YouTube}}'s {{Recommendation Systems}}},
  author = {Chandio, Sarmad and Dar, Daniyal Pirwani and Nithyanand, Rishab},
  date = {2023},
  journaltitle = {arXiv:2303.03445},
  doi = {10.48550/arXiv.2303.03445}
}

@online{chowdhuryWhatAudit,
  title = {What’s an Audit?},
  author = {Chowdhury, Rumman},
  date = {2023-05-29},
  url = {https://www.get-parity.com/raiblog/whats-an-audit},
  urldate = {2023-12-21},
  organization = {Parity Consulting}
}

@incollection{christofiDataProtectionControl2022,
  title = {Data Protection, Control and Participation beyond Consent - {{Seeking}} the Views of Data Subjects in Data Protection Impact Assessments},
  booktitle = {Research {{Handbook}} on {{EU Data Protection Law}}},
  author = {Christofi, Athena and Breuer, Jonas and Wauters, Ellen and Valcke, Peggy and Pierson, Jo},
  date = {2022-04-22},
  pages = {503--529},
  publisher = {Edward Elgar Publishing},
  doi = {10.4337/9781800371682.00029}
}

@article{cobbeArtificialIntelligenceService2021,
  title = {Artificial Intelligence as a Service: {{Legal}} Responsibilities, Liabilities, and Policy Challenges},
  shorttitle = {Artificial Intelligence as a Service},
  author = {Cobbe, Jennifer and Singh, Jatinder},
  date = {2021},
  journaltitle = {Computer Law \& Security Review},
  volume = {42},
  pages = {105573},
  doi = {10/gmq8jm}
}

@inproceedings{cobbeUnderstandingAccountabilityAlgorithmic2023,
  title = {Understanding Accountability in Algorithmic Supply Chains},
  booktitle = {Proceedings of the 2023 {{ACM Conference}} on {{Fairness}}, {{Accountability}}, and {{Transparency}}},
  author = {Cobbe, Jennifer and Veale, Michael and Singh, Jatinder},
  date = {2023-06-12},
  series = {{{FAccT}} '23},
  pages = {1186--1197},
  publisher = {Association for Computing Machinery},
  location = {New York, NY, USA},
  doi = {gsb98p},
}

@inproceedings{costanzachockWhoAuditsAuditors2022,
  title = {Who {{Audits}} the {{Auditors}}? {{Recommendations}} from a Field Scan of the Algorithmic Auditing Ecosystem},
  author = {Costanza-Chock, Sasha and Raji, Inioluwa Deborah and Buolamwini, Joy},
  date = {2022},
  pages = {1571--1583},
  booktitle = {Proceedings of the 2022 {{ACM Conference}} on {{Fairness}}, {{Accountability}}, and {{Transparency}}},
  publisher = {Association for Computing Machinery},
  series = {{FAccT '22}},
  doi = {10.1145/3531146.3533213}
}

@inproceedings{damourFairnessNotStatic2020,
  title = {Fairness Is Not Static: Deeper Understanding of Long Term Fairness via Simulation Studies},
  shorttitle = {Fairness Is Not Static},
  booktitle = {Proceedings of the 2020 {{Conference}} on {{Fairness}}, {{Accountability}}, and {{Transparency}}},
  author = {D'Amour, Alexander and Srinivasan, Hansa and Atwood, James and Baljekar, Pallavi and Sculley, D. and Halpern, Yoni},
  date = {2020-01-27},
  series = {{{FAT}}* '20},
  pages = {525--534},
  publisher = {Association for Computing Machinery},
  location = {New York, NY, USA},
  doi = {10.1145/3351095.3372878},
}

@online{dariusImplementingDataAccess2023,
  title = {Implementing {{Data Access}} of the {{Digital Services Act}}: {{Collaboration}} of {{European Digital Service Coordinators}} and {{Researchers}} in {{Building Strong Oversight}} over {{Social Media Platforms}}},
  author = {Darius, Philipp and Stockmann, Daniela and Bryson, Joanna and Cingolani, Luciana and Griffin, Rachel and Hammerschmid, Gerhard and Kupi, Maximilian and Mones, Haytham and Munzert, Simon and Riordan, Rónán},
  date = {2023},
  url = {https://perma.cc/P2MM-QWUP},
  organisation = {Hertie School of Governance}
}

@online{DeloitteAIInstitute,
  title = {Deloitte {{AI Institute Teams With Chatterbox Labs}} to {{Ensure Ethical Application}} of {{AI}}},
  url = {https://perma.cc/DC8D-QAS5},
  organization = {{PR Newswire}},
  author = {{Deloitte Consulting LLP}},
  date = {2021-03-21}
}

@online{DeloitteIntroducesTrustworthy,
  title = {{Deloitte Introduces Trustworthy AI Framework to Guide Organizations in Ethical Application of Technology in the Age of With}},
  url = {https://perma.cc/RZ3Q-UD47},
  organization = {{PR Newswire}},
  author = {{Deloitte Consulting LLP}},
  date = {2020-08-26}
}

@article{dergachevaImprovingDataAccess2023,
  journaltitle = {SSRN Scholarly Paper},
  title = {Improving {{Data Access}} for {{Researchers}} in the {{Digital Services Act}}},
  author = {Dergacheva, Daria and Katzenbach, Christian and Schwemer, Sebastian Felix and Quintais, João Pedro},
  date = {2023-06-01},
  doi = {10.2139/ssrn.4465846},
}

@online{DigitalServicesAct,
  title = {Digital {{Services Act Audit}} - {{Solution}}},
  url = {https://www.holisticai.com/digital-services-act-audit},
  urldate = {2024-01-16},
  author = {{Holistic AI}}
}

@article{dimaggioIronCageRevisited1983,
  title = {The {{Iron Cage Revisited}}: {{Institutional Isomorphism}} and {{Collective Rationality}} in {{Organizational Fields}}},
  shorttitle = {The {{Iron Cage Revisited}}},
  author = {DiMaggio, Paul J. and Powell, Walter W.},
  date = {1983},
  journaltitle = {American Sociological Review},
  volume = {48},
  number = {2},
  pages = {147},
  doi = {10.2307/2095101}
}

@incollection{edwardsGreatPowerComes2019,
  title = {‘{{With Great Power Comes Great Responsibility}}?’: {{The Rise}} of {{Platform Liability}}},
  booktitle = {Law, {{Policy}}, and the {{Internet}}},
  author = {Edwards, Lilian},
  editor = {Edwards, Lilian},
  date = {2019},
  pages = {253--289},
  publisher = {Hart Publishing},
}

@book{elderAuditingAssuranceServices2020,
  title = {Auditing and {{Assurance Services}}: {{International Perspectives}}},
  shorttitle = {Auditing and {{Assurance Services}}},
  author = {Elder, Randal J. and Beasley, Mark S. and Hogan, Chris E. and Arens, Alvin A.},
  date = {2020},
  edition = {17},
  publisher = {Pearson}
}

@online{EqualityImpactAssessment,
  title = {Equality {{Impact Assessment}}: {{A}} Practical Tool to Identify Discrimination},
  author = {South Wales Police},
  url = {https://perma.cc/E7AU-VFNJ},
  date = {2021-08-10}
}

@online{EUAIAct,
  title = {{{EU AI Act Readiness}} - {{Solutions}}},
  url = {https://www.holisticai.com/eu-ai-act-readiness},
  urldate = {2024-01-16},
  author = {{Holistic AI}}
}

@online{EUDigitalServices,
  title = {The {{EU}}’s {{Digital Services Act}}},
  url = {https://commission.europa.eu/strategy-and-policy/priorities-2019-2024/europe-fit-digital-age/digital-services-act_en},
  urldate = {2024-01-19},
  author = {{European Commission}}
}

@online{europeanparliamentEuropeanHealthData2023,
  title = {European {{Health Data Space Regulation}} ({{Amendments}} Adopted by the {{European Parliament}})},
  author = {{European Parliament}},
  date = {2023-12-13},
  url = {https://www.europarl.europa.eu/doceo/document/TA-9-2023-0462_EN.html}
}

@report{EY,
  title = {{{EY}}}
}

@online{EYTrustedAI,
  title = {{{EY Trusted AI Platform}}},
  url = {https://www.ey.com/en_gl/consulting/trusted-ai-platform},
  author = {{Ernst and Young}}
}

@online{federaltradecommissionCombattingOnlineHarms2022,
  title = {Combatting {{Online Harms Through Innovation}}},
  author = {{Federal Trade Commission}},
  date = {2022-06-16},
  url = {https://perma.cc/2S5Y-WQKX}
}

@online{FiddlerAIAI,
  title = {{{AI Observability}}, {{ML Model Monitoring}}, and {{Explainable AI}}},
  url = {https://www.fiddler.ai/},
  urldate = {2024-01-16},
  author = {{Fiddler AI}}
}

@online{golbinAlgorithmicImpactAssessments,
  title = {Algorithmic Impact Assessments: {{What}} Are They and Why Do You Need Them?},
  shorttitle = {Algorithmic Impact Assessments},
  author = {Golbin, Ilana},
  url = {https://perma.cc/6CFY-MYNZ},
  urldate = {2023-12-13},
  organization = {PwC},
  date = {2021-10-08}
}

@article{goodmanALGORITHMICAUDITINGCHASING2023,
  title = {{{Algorithmic Auditing}}: {{Chasing AI Accountability}}},
  author = {Goodman, Ellen P and Tr\'{e}hu, Julia},
  date = {2023},
  journaltitle = {Santa Clara High Technology Law Journal},
  volume = {39},
  number = {3},
  pages = {289}
}

@article{goodwinProfessionalVision1994,
  title = {Professional {{Vision}}},
  author = {Goodwin, Charles},
  date = {1994},
  journaltitle = {American Anthropologist},
  volume = {96},
  number = {3},
  pages = {606--633},
  publisher = {[American Anthropological Association, Wiley]},
  url = {https://www.jstor.org/stable/682303},
  urldate = {2024-01-12}
}

@online{googleSubmissionsGoogleResponse2023,
  title = {Submissions by {{Google}} in Response to the {{European Commission}}’s Request for Feedback on the Draft {{DRPA}}},
  author = {{Google}},
  date = {2023-06-02},
  url = {https://ec.europa.eu/info/law/better-regulation/have-your-say/initiatives/13626-Digital-Services-Act-conducting-independent-audits/F3424072_en},
  urldate = {2023-12-08}
}

@article{gorwaAlgorithmicContentModeration2020,
  title = {Algorithmic Content Moderation: {{Technical}} and Political Challenges in the Automation of Platform Governance},
  shorttitle = {Algorithmic Content Moderation},
  author = {Gorwa, Robert and Binns, Reuben and Katzenbach, Christian},
  date = {2020},
  journaltitle = {Big Data \& Society},
  volume = {7},
  number = {1},
  pages = {2053951719897945},
  doi = {10/ggsfrk}
}

@article{gorwaModeratingModelMarketplaces2024,
  title = {Moderating {{Model Marketplaces}}: {{Platform Governance Puzzles}} for {{AI Intermediaries}}},
  shorttitle = {Moderating {{Model Marketplaces}}},
  author = {Gorwa, Robert and Veale, Michael},
  date = {2024},
  journaltitle = {Law, Innovation and Technology},
  volume = {16},
  number = {2},
  doi = {10.31235/osf.io/6dfk3}
}

@book{gorwaPoliticsPlatformRegulation2024,
  title = {The {{Politics}} of {{Platform Regulation}}: {{How Governments Shape Online Content Moderation}}},
  author = {Gorwa, Robert},
  date = {2024},
  publisher = {Oxford University Press},
  location = {Oxford, UK}
}

@online{governmentofcanadaSummaryRegulatoryPowers2022,
  title = {Summary: {{Regulatory Powers}}},
  shorttitle = {Summary of {{Session Four}}},
  author = {{Government of Canada}},
  date = {2022-05-16},
  url = {https://www.canada.ca/en/canadian-heritage/campaigns/harmful-online-content/summary-session-four.html}
}

@online{balaynDebiasingRegulatingAI2021,
  title = {Beyond Debiasing: {{Regulating AI}} and Its Inequalities},
  author = {Balayn, Agathe and Gürses, Seda},
  date = {2021},
  url = {https://perma.cc/4UAV-3UFB},
  urldate = {2021-09-26},
  organization = {European Digital Rights (EDRi)}
}

@book{hartmannEvolutionIntermediaryInstitutions2015,
  title = {The {{Evolution}} of {{Intermediary Institutions}} in {{Europe}}},
  editor = {Hartmann, Eva and Kjaer, Poul F.},
  date = {2015},
  publisher = {Palgrave Macmillan UK},
  location = {London},
  doi = {10.1057/9781137484529},
}

@online{hmgovernmentImpactAssessmentOnline2022,
  title = {Impact {{Assessment}}: The {{Online Safety Bill}}},
  author = {{HM Government}},
  date = {2022-03-17},
  url = {https://publications.parliament.uk/pa/bills/cbill/58-02/0285/210285en.pdf}
}

@book{hmtreasuryAquaBookGuidance2015,
  title = {The {{Aqua Book}}: Guidance on Producing Quality Analysis for Government},
  author = {{HM Treasury}},
  date = {2015},
  publisher = {HM Government},
  location = {London}
}

@online{hobbhahnAnnouncingApolloResearch,
  title = {Announcing {{Apollo Research}}},
  author = {Hobbhahn, Marius},
  url = {https://forum.effectivealtruism.org/posts/ysC6crBKhDBGZfob3/announcing-apollo-research},
  organization = {Effective Altruism Forum},
  date = {2021-03-31}
}

@inproceedings{holsteinImprovingFairnessMachine2019,
  title = {Improving {{Fairness}} in {{Machine Learning Systems}}: {{What Do Industry Practitioners Need}}?},
  shorttitle = {Improving {{Fairness}} in {{Machine Learning Systems}}},
  booktitle = {Proceedings of the 2019 {{CHI Conference}} on {{Human Factors}} in {{Computing Systems}}},
  author = {Holstein, Kenneth and Wortman Vaughan, Jennifer and Daum\'{e}, Hal and Dudik, Miro and Wallach, Hanna},
  date = {2019-05-02},
  series = {{{CHI}} '19},
  pages = {1--16},
  publisher = {Association for Computing Machinery},
  location = {New York, NY, USA},
  doi = {10.1145/3290605.3300830},
}

@article{imanaHavingYourPrivacy2023,
  title = {Having Your {{Privacy Cake}} and {{Eating}} It {{Too}}: {{Platform-supported Auditing}} of {{Social Media Algorithms}} for {{Public Interest}}},
  author = {Imana, Basileal and Korolova, Aleksandra and Heidemann, John},
  date = {2023},
  journaltitle = {Proceedings of the ACM on Human-Computer Interaction},
  volume = {7},
  pages = {1--33},
  publisher = {ACM New York, NY, USA},
  issue = {CSCW1}
}

@online{IntroductionConformityAssessment2022,
  title = {Introduction to the {{Conformity Assessment}} under the {{Draft EU AI Act}}, and How It {{Compares}} to {{DPIAs}}},
  date = {2022-08-12},
  url = {https://fpf.org/blog/introduction-to-the-conformity-assessment-under-the-draft-eu-ai-act-and-how-it-compares-to-dpias/},
  organization = {Future of Privacy Forum},
  author = {Demetzou, Katarina}
}

@article{vealeDemystifyingDraftEU2021,
  title = {Demystifying the {{Draft EU Artificial Intelligence Act}}},
  author = {Veale, Michael and Zuiderveen Borgesius, Frederik},
  date = {2021},
  journaltitle = {Computer Law Review International},
  volume = {22},
  number = {4},
  pages = {97--112},
  doi = {10/gns2s9}
}

@article{iosifidisPublicSphereSocial2011,
  title = {The Public Sphere, Social Networks and Public Service Media},
  author = {Iosifidis, Petros},
  date = {2011},
  journaltitle = {Information, Communication \& Society},
  volume = {14},
  number = {5},
  pages = {619--637},
  publisher = {Taylor \& Francis}
}

@online{kayeAIRegulationCoulda,
  title = {{{AI}} Regulation Could Spur New Auditing Services for Tech Companies},
  author = {Kaye, Kate},
  url = {https://www.protocol.com/enterprise/ai-audit-2022},
  urldate = {2024-01-20},
  organization = {Protocol}
}

@online{abelsonBugsOurPockets2021a,
  title = {Bugs in Our {{Pockets}}: {{The Risks}} of {{Client-Side Scanning}}},
  shorttitle = {Bugs in Our {{Pockets}}},
  author = {Abelson, Hal and Anderson, Ross and Bellovin, Steven M. and Benaloh, Josh and Blaze, Matt and Callas, Jon and Diffie, Whitfield and Landau, Susan and Neumann, Peter G. and Rivest, Ronald L. and Schiller, Jeffrey I. and Schneier, Bruce and Teague, Vanessa and Troncoso, Carmela},
  date = {2021-10-14},
  number = {2110.07450v1},
  eprint = {2110.07450v1},
  eprinttype = {arxiv},
  url = {https://arxiv.org/abs/2110.07450v1},
  urldate = {2021-12-22},
}

@article{koeneckeRacialDisparitiesAutomated2020,
  title = {Racial Disparities in Automated Speech Recognition},
  author = {Koenecke, Allison and Nam, Andrew and Lake, Emily and Nudell, Joe and Quartey, Minnie and Mengesha, Zion and Toups, Connor and Rickford, John R. and Jurafsky, Dan and Goel, Sharad},
  date = {2020-04-07},
  journaltitle = {Proceedings of the National Academy of Sciences},
  volume = {117},
  number = {14},
  pages = {7684--7689},
  publisher = {Proceedings of the National Academy of Sciences},
  doi = {10.1073/pnas.1915768117}
}

@article{lauxTamingFewPlatform2021,
  title = {Taming the Few: {{Platform}} Regulation, Independent Audits, and the Risks of Capture Created by the {{DMA}} and {{DSA}}},
  author = {Laux, Johann and Wachter, Sandra and Mittelstadt, Brent},
  date = {2021},
  journaltitle = {Computer Law \& Security Review},
  volume = {43},
  pages = {105613},
  publisher = {Elsevier}
}

@incollection{leerssenScrapingEuropeLaw2023,
  title = {Scraping {{By}}? {{Europe}}'s Law and Policy on Social Media Research Access},
  shorttitle = {Scraping {{By}}?},
  booktitle = {Challenges and Perspectives of Hate Speech Research},
  author = {Leerssen, Paddy and Heldt, Amélie P. and Kettemann, Matthias C.},
  editor = {Strippel, Christian and Paasch-Colberg, Sünje and Emmer, Martin and Trebbe, Joachim},
  date = {2023},
  series = {Digital {{Communication Research}}},
  volume = {12},
  pages = {405--425},
  location = {Berlin},
  doi = {10.48541/dcr.v12.24},
}

@thesis{leerssenSeeingWhatOthers2023,
  type = {phdthesis},
  title = {Seeing What Others Are Seeing: {{Studies}} in the Regulation of Transparency for Social Media Recommender Systems},
  author = {Leerssen, Paddy},
  date = {2023},
  institution = {University of Amsterdam}
}

@legislation{LocalLawInt2023,
  title = {Local {{Law Int}}. {{No}}. 1894-{{A}} ({{Automated Employment Decision Tools Law}} ("{{AEDT}}")), {{New York City}}},
  date = {2023}
}

@online{lomasSecurityResearchersLatest2023,
  title = {Security Researchers Latest to Blast {{UK}}'s {{Online Safety Bill}} as Encryption Risk},
  author = {Lomas, Natasha},
  date = {2023-07-05T11:04:03+00:00},
  url = {https://techcrunch.com/2023/07/05/uk-online-safety-bill-risks-e2ee/},
  urldate = {2024-01-18},
  organization = {TechCrunch}
}

@article{macsithighRoadResponsibilitiesNew2020,
  title = {The {{Road}} to {{Responsibilities}}: {{New Attitudes Towards Internet Intermediaries}}},
  author = {Mac Síthigh, Daithí},
  date = {2020},
  journaltitle = {Information \& Communications Technology Law},
  volume = {29},
  number = {1},
  pages = {1--21},
  doi = {10/ggkhjn}
}

@inproceedings{madaioCoDesigningChecklistsUnderstand2020,
  title = {Co-{{Designing Checklists}} to {{Understand Organizational Challenges}} and {{Opportunities}} around {{Fairness}} in {{AI}}},
  booktitle = {Proceedings of the 2020 {{CHI Conference}} on {{Human Factors}} in {{Computing Systems}}},
  author = {Madaio, Michael A. and Stark, Luke and Wortman Vaughan, Jennifer and Wallach, Hanna},
  date = {2020-04-23},
  series = {{{CHI}} '20},
  pages = {1--14},
  publisher = {Association for Computing Machinery},
  location = {New York, NY, USA},
  doi = {10.1145/3313831.3376445},
}

@inproceedings{maffeyMLTEingModelsNegotiating2023,
  title = {{{MLTEing Models}}: {{Negotiating}}, {{Evaluating}}, and {{Documenting Model}} and {{System Qualities}}},
  shorttitle = {{{MLTEing Models}}},
  booktitle = {2023 {{IEEE}}/{{ACM}} 45th {{International Conference}} on {{Software Engineering}}: {{New Ideas}} and {{Emerging Results}} ({{ICSE-NIER}})},
  author = {Maffey, Katherine R. and Dotterrer, Kyle and Niemann, Jennifer and Cruickshank, Iain and Lewis, Grace A. and Kästner, Christian},
  date = {2023-05},
  pages = {31--36},
  doi = {10.1109/ICSE-NIER58687.2023.00012},
  eventtitle = {2023 {{IEEE}}/{{ACM}} 45th {{International Conference}} on {{Software Engineering}}: {{New Ideas}} and {{Emerging Results}} ({{ICSE-NIER}})}
}

@book{kjaerLawPoliticalEconomy2020,
  title = {The {{Law}} of {{Political Economy}}: {{Transformation}} in the {{Function}} of {{Law}}},
  shorttitle = {The {{Law}} of {{Political Economy}}},
  editor = {Kjaer, Poul F.},
  date = {2020},
  publisher = {Cambridge University Press},
  location = {Cambridge},
  doi = {10.1017/9781108675635},
}

@article{sobelreadReimaginingUnimaginableLaw2020,
  title = {Reimagining the {{Unimaginable}}: {{Law}} and the {{Ongoing Transformation}} of {{Global Value Chains}} into {{Integrated Legal Entities}}},
  shorttitle = {Reimagining the {{Unimaginable}}},
  author = {Sobel-Read, Kevin B.},
  date = {2020-04-07},
  journaltitle = {European Review of Contract Law},
  volume = {16},
  number = {1},
  pages = {160--185},
  doi = {10.1515/ercl-2020-0009}
}

@online{markdangelo2023,
  author = {Mark Dangelo},
  date = {2023-10-25},
  url = {https://www.thomsonreuters.com/en-us/posts/technology/auditing-ai-transparency/},
  title = {Auditing {AI: T}he emerging battlefield of transparency and assessment},
  organization = {Thompson Reuters}
}

@article{matusCertificationSystemsMachine2022,
  title = {Certification Systems for Machine Learning: {{Lessons}} from Sustainability},
  shorttitle = {Certification Systems for Machine Learning},
  author = {Matus, Kira J. M. and Veale, Michael},
  date = {2022},
  journaltitle = {Regulation \& Governance},
  volume = {16},
  number = {1},
  pages = {177--196},
  doi = {10.1111/rego.12417}
}

@article{messmerAuditingRecommenderSystems2023,
  title = {Auditing {{Recommender Systems--Putting}} the {{DSA}} into Practice with a Risk-Scenario-Based Approach},
  author = {Meßmer, Anna-Katharina and Degeling, Martin},
  date = {2023},
  journaltitle = {Stiftung Neue Verantwortung},
  doi = {10.48550/arXiv.2302.04556}
}

@article{metcalfOwningEthicsCorporate,
  title = {Owning {{Ethics}}: {{Corporate Logics}}, {{Silicon Valley}}, and the {{Institutionalization}} of {{Ethics}}},
  shorttitle = {Owning {{Ethics}}},
  author = {Metcalf, Jacob and Moss, Emanuel and Boyd, Danah},
  date = {2019},
  journaltitle = {Social Research: An International Quarterly},
  volume = {86},
  number = {2},
  pages = {449--476},
  doi = {10.1353/sor.2019.0022}
}

@article{mitchellAlgorithmicFairnessChoices2021,
  title = {Algorithmic {{Fairness}}: {{Choices}}, {{Assumptions}}, and {{Definitions}}},
  shorttitle = {Algorithmic {{Fairness}}},
  author = {Mitchell, Shira and Potash, Eric and Barocas, Solon and D'Amour, Alexander and Lum, Kristian},
  date = {2021-03-07},
  journaltitle = {Annu. Rev. Stat. Appl.},
  volume = {8},
  number = {1},
  pages = {141--163},
  doi = {10.1146/annurev-statistics-042720-125902}
}

@inproceedings{naharCollaborationChallengesBuilding2021,
  title = {Collaboration Challenges in Building {{ML-enabled}} Systems: Communication, Documentation, Engineering, and Process},
  shorttitle = {Collaboration Challenges in Building {{ML-enabled}} Systems},
  booktitle = {Proceedings of the 44th {{International Conference}} on {{Software Engineering}}},
  author = {Nahar, Nadia and Zhou, Shurui and Lewis, Grace and Kästner, Christian},
  date = {2022-07-05},
  series = {{{ICSE}} '22},
  pages = {413--425},
  publisher = {Association for Computing Machinery},
  location = {New York, NY, USA},
  doi = {10.1145/3510003.3510209}
}

@inproceedings{nanniniExplainabilityAIPolicies2023,
  title = {Explainability in {{AI Policies}}: {{A Critical Review}} of {{Communications}}, {{Reports}}, {{Regulations}}, and {{Standards}} in the {{EU}}, {{US}}, and {{UK}}},
  shorttitle = {Explainability in {{AI Policies}}},
  booktitle = {Proceedings of the 2023 {{ACM Conference}} on {{Fairness}}, {{Accountability}}, and {{Transparency}}},
  author = {Nannini, Luca and Balayn, Agathe and Smith, Adam Leon},
  date = {2023-06-12},
  series = {{{FAccT}} '23},
  pages = {1198--1212},
  publisher = {Association for Computing Machinery},
  location = {New York, NY, USA},
  doi = {10.1145/3593013.3594074},
}

@inproceedings{norvalNavigatingAuditLandscape2023,
  title = {Navigating the {{Audit Landscape}}: {{A Framework}} for {{Developing Transparent}} and {{Auditable XR}}},
  shorttitle = {Navigating the {{Audit Landscape}}},
  booktitle = {Proceedings of the 2023 {{ACM Conference}} on {{Fairness}}, {{Accountability}}, and {{Transparency}}},
  author = {Norval, Chris and Cloete, Richard and Singh, Jatinder},
  date = {2023-06-12},
  series = {{{FAccT}} '23},
  pages = {1418--1431},
  publisher = {Association for Computing Machinery},
  location = {New York, NY, USA},
  doi = {10.1145/3593013.3594090},
}

@online{ofcomProtectingPeopleIllegal2023b,
  title = {Protecting People from Illegal Harms Online - {{Summary}} of Each Chapter},
  author = {{Ofcom}},
  date = {2023-11-09},
  url = {https://www.ofcom.org.uk/__data/assets/pdf_file/0030/270948/chapter-summaries-illegal-harms-consultation.pdf}
}

@article{zhuHowLeadAuditor2024,
  title = {How Do Lead Auditor Instructions Influence Component Auditors' Evidence Collection Decisions? {{The}} Joint Influence of Construal Interpretations and Responsibility},
  shorttitle = {How Do Lead Auditor Instructions Influence Component Auditors' Evidence Collection Decisions?},
  author = {Zhu, Skye and Phang, Soon-Yeow},
  date = {2024},
  journaltitle = {Contemporary Accounting Research},
  volume = {41},
  number = {1},
  pages = {591--619},
  doi = {10.1111/1911-3846.12911}
}

@article{sunderlandMultinationalGroupAudits2017,
  title = {Multinational {{Group Audits}}: {{Problems Faced}} in {{Practice}} and {{Opportunities}} for {{Research}}},
  shorttitle = {Multinational {{Group Audits}}},
  author = {Sunderland, Dan and Trompeter, Gregory M.},
  date = {2017},
  journaltitle = {Auditing: A Journal of Practice \& Theory},
  volume = {36},
  number = {3},
  pages = {159--183},
  doi = {10.2308/ajpt-51667}
}

@article{downeyChallengingGlobalGroup2021,
  title = {Challenging {{Global Group Audits}}: {{The Perspective}} of {{US Group Audit Leads}}},
  shorttitle = {Challenging {{Global Group Audits}}},
  author = {Downey, Denise Hanes and Westermann, Kimberly D.},
  date = {2021},
  journaltitle = {Contemporary Accounting Research},
  volume = {38},
  number = {2},
  pages = {1395--1433},
  doi = {10.1111/1911-3846.12648}
}

@inproceedings{okoloMakingAIExplainable2022,
  title = {Making {{AI Explainable}} in the {{Global South}}: {{A Systematic Review}}},
  shorttitle = {Making {{AI Explainable}} in the {{Global South}}},
  booktitle = {Proceedings of the 5th {{ACM SIGCAS}}/{{SIGCHI Conference}} on {{Computing}} and {{Sustainable Societies}}},
  author = {Okolo, Chinasa T. and Dell, Nicola and Vashistha, Aditya},
  date = {2022},
  series = {{{COMPASS}} '22},
  pages = {439--452},
  publisher = {Association for Computing Machinery},
  location = {New York, NY, USA},
  doi = {10.1145/3530190.3534802},
}

@online{hmgovernmentFuturesToolkitTools2017,
  title = {The {{Futures Toolkit}}: {{Tools}} for {{Futures Thinking}} and {{Foresight}} across {{UK Government}}},
  author = {{HM Government}},
  date = {2017},
  url = {https://assets.publishing.service.gov.uk/media/5a821fdee5274a2e8ab579ef/futures-toolkit-edition-1.pdf}
}

@online{OnlineSafetyBill2023,
  title = {Online {{Safety Bill}}, {{Hansard}} ({{HL}} Vol 831, Col 385)},
  date = {2023-06-22},
  url = {https://hansard.parliament.uk/lords/2023-06-22/debates/C956350F-D70E-4409-8A4D-4ED05488C6DE/OnlineSafetyBill},
  urldate = {2024-01-18},
  organization = {Hansard (HL vol 831, col 385)}
}

@online{OpenLetterSecurity,
  title = {Open {{Letter}} from {{Security}} and {{Privacy Researchers}} in Relation to the {{Online Safety Bill}}},
  url = {https://haddadi.github.io/UKOSBOpenletter.pdf}
}

@inproceedings{pangAuditingCrossCulturalConsistency2023,
  title = {Auditing {{Cross-Cultural Consistency}} of {{Human-Annotated Labels}} for {{Recommendation Systems}}},
  booktitle = {Proceedings of the 2023 {{ACM Conference}} on {{Fairness}}, {{Accountability}}, and {{Transparency}}},
  author = {Pang, Rock Yuren and Cenatempo, Jack and Graham, Franklyn and Kuehn, Bridgette and Whisenant, Maddy and Botchway, Portia and Stone Perez, Katie and Koenecke, Allison},
  date = {2023-06-12},
  series = {{{FAccT}} '23},
  pages = {1531--1552},
  publisher = {Association for Computing Machinery},
  location = {New York, NY, USA},
  doi = {10.1145/3593013.3594098},
}

@inproceedings{grovesAuditingWorkExploring2024,
  title = {Auditing {{Work}}: {{Exploring}} the {{New York City}} Algorithmic Bias Audit Regime},
  shorttitle = {Auditing {{Work}}},
  booktitle = {Proceedings of the 2024 {{Conference}} on {{Fairness}}, {{Accountability}}, and {{Transparency}} in {{Algorithmic Systems}} ({{FAccT}}'24)},
  author = {Groves, Lara and Metcalf, Jacob and Kennedy, Alayna and Vecchione, Briana and Strait, Andrew},
  date = {2024},
  publisher = {ACM},
  location = {Rio de Janeiro, Brazil},
  doi = {10.48550/arXiv.2402.08101}
}

@book{pbluncertainty,
  title = {Guidance for Uncertainty Assessment and Communication},
  author = {Petersen, Arthur C and Janssen, P H M and family=Sluijs, given=J P, prefix=van der, useprefix=true and Risbet, J S and Ravetz, J R and Arjan Wardekker, J and Martison Hughes, H},
  date = {2013},
  publisher = {PBL Netherlands Environmental Assessment Bureau},
  location = {The Hague, NL},
  url = {https://perma.cc/N9FJ-RGXU}
}

@online{peersmanFrameworkEvaluatingCSAM2023,
  title = {Towards a {{Framework}} for {{Evaluating CSAM Prevention}} and {{Detection Tools}} in the {{Context}} of {{End-to-end}} Encryption {{Environments}}: A {{Case Study}}},
  author = {Peersman, Claudia and Llanos, Jos\'{e} Tomas and May-Chahal, Corinne and McConville, Ryan and Das Chowdhury, Partha and De Cristofaro, Emiliano},
  date = {2023-02},
  url = {https://www.rephrain.ac.uk/safety-tech-challenge-fund/},
  urldate = {2024-01-18},
  organization = {REPHRAIN}
}

@article{poonNewDealInstitutions2009,
  title = {From New Deal Institutions to Capital Markets: {{Commercial}} Consumer Risk Scores and the Making of Subprime Mortgage Finance},
  shorttitle = {From New Deal Institutions to Capital Markets},
  author = {Poon, Martha},
  date = {2009},
  journaltitle = {Accounting, Organizations and Society},
  shortjournal = {Accounting, Organizations and Society},
  volume = {34},
  number = {5},
  pages = {654--674},
  doi = {10/cm8g3x}
}

@article{poonScorecardsDevicesConsumer2007,
  title = {Scorecards as {{Devices}} for {{Consumer Credit}}: {{The Case}} of {{Fair}}, {{Isaac}} \& {{Company Incorporated}}},
  shorttitle = {Scorecards as {{Devices}} for {{Consumer Credit}}},
  author = {Poon, Martha},
  date = {2007},
  journaltitle = {The Sociological Review},
  shortjournal = {The Sociological Review},
  volume = {55},
  pages = {284--306},
  doi = {10/bbzb34},
  issue = {2\_suppl}
}

@article{phiriSocialNetworksCorruption2019,
  title = {Social Networks, Corruption and Institutions of Accounting, Auditing and Accountability},
  author = {Phiri, Joseph and Guven-Uslu, Pinar},
  date = {2019},
  journaltitle = {Accounting, Auditing \& Accountability Journal},
  volume = {32},
  number = {2},
  pages = {508--530},
  publisher = {Emerald Publishing Limited}
}

@report{PolicyDocumentOvert,
  title = {Policy {{Document}} for the Overt Deployment of {{Live Facial Recognition}} ({{LFR}}) {{Technology}}},
  url = {https://perma.cc/H5NF-D9QT},
  author = {{South Wales Police}},
  date = {2023-06-29}
}

@inproceedings{rajiAlgorithmicAuditsActual2022,
  title = {From {{Algorithmic Audits}} to {{Actual Accountability}}: {{Overcoming Practical Roadblocks}} on the {{Path}} to {{Meaningful Audit Interventions}} for {{AI Governance}}},
  shorttitle = {From {{Algorithmic Audits}} to {{Actual Accountability}}},
  booktitle = {Proceedings of the 2022 {{AAAI}}/{{ACM Conference}} on {{AI}}, {{Ethics}}, and {{Society}}},
  author = {Raji, Inioluwa Deborah},
  date = {2022-07-27},
  series = {{{AIES}} '22},
  pages = {5},
  publisher = {Association for Computing Machinery},
  location = {New York, NY, USA},
  doi = {10.1145/3514094.3539566},
}

@inproceedings{rajiClosingAIAccountability2020,
  title = {Closing the {{AI}} Accountability Gap: Defining an End-to-End Framework for Internal Algorithmic Auditing},
  shorttitle = {Closing the {{AI}} Accountability Gap},
  booktitle = {Proceedings of the 2020 {{Conference}} on {{Fairness}}, {{Accountability}}, and {{Transparency}}},
  author = {Raji, Inioluwa Deborah and Smart, Andrew and White, Rebecca N. and Mitchell, Margaret and Gebru, Timnit and Hutchinson, Ben and Smith-Loud, Jamila and Theron, Daniel and Barnes, Parker},
  date = {2020-01-27},
  series = {{{FAT}}* '20},
  pages = {33--44},
  publisher = {Association for Computing Machinery},
  location = {New York, NY, USA},
  doi = {10.1145/3351095.3372873},
}

@online{internetsafetygreen,
author = {{HM Government}},
month = oct,
year = 2017, 
title = {{Internet Safety Strategy -- Green paper}},
url = {https://www.gov.uk/government/consultations/internet-safety-strategy-green-paper}
}

@online{woodsDraftOnlineHarm2019,
  title = {Draft {{Online Harm Reduction Bill}}},
  author = {Woods, Lorna and Perrin, Will and Walsh, Maeve},
  date = {2019},
  url = {https://carnegieuktrust.org.uk/publications/draft-online-harm-bill/},
  urldate = {2024-04-01},
  organization = {Carnegie UK Trust}
}

@book{marsdenInternetCoRegulationEuropean2011,
  title = {Internet {{Co-Regulation}}: {{European Law}}, {{Regulatory Governance}} and {{Legitimacy}} in {{Cyberspace}}},
  shorttitle = {Internet {{Co-Regulation}}},
  author = {Marsden, Christopher T.},
  date = {2011},
  publisher = {Cambridge University Press},
  urldate = {2020},
}

@incollection{woodsObligingPlatformsAccept2021,
  title = {Obliging {{Platforms}} to {{Accept}} a {{Duty}} of {{Care}}},
  booktitle = {Regulating {{Big Tech}}},
  author = {Woods, Lorna and Perrin, Will},
  editor = {Moore, Martin and Tambini, Damian},
  date = {2021},
  pages = {93--109},
  publisher = {Oxford University Press},
  location = {Oxford},
  doi = {10.1093/oso/9780197616093.003.0006},
}

@online{rajiItTimeDevelop2022,
  title = {It’s {{Time}} to {{Develop}} the {{Tools We Need}} to {{Hold Algorithms Accountable}}},
  author = {Raji, Inioluwa Deborah},
  date = {2022-02-02},
  url = {https://foundation.mozilla.org/en/blog/its-time-to-develop-the-tools-we-need-to-hold-algorithms-accountable/},
  urldate = {2024-01-12},
  organization = {Mozilla Foundation}
}

@online{rajiSavingFaceInvestigating2020,
  title = {Saving {{Face}}: {{Investigating}} the {{Ethical Concerns}} of {{Facial Recognition Auditing}}},
  shorttitle = {Saving {{Face}}},
  author = {Raji, Inioluwa Deborah and Gebru, Timnit and Mitchell, Margaret and Buolamwini, Joy and Lee, Joonseok and Denton, Emily},
  date = {2020-01-03},
  eprint = {2001.00964},
  eprinttype = {arxiv},
  eprintclass = {cs},
  doi = {10.48550/arXiv.2001.00964},
  pubstate = {preprint}
}

@article{rakovaWhereResponsibleAI2021,
  title = {Where {{Responsible AI}} Meets {{Reality}}: {{Practitioner Perspectives}} on {{Enablers}} for {{Shifting Organizational Practices}}},
  shorttitle = {Where {{Responsible AI}} Meets {{Reality}}},
  author = {Rakova, Bogdana and Yang, Jingying and Cramer, Henriette and Chowdhury, Rumman},
  date = {2021-04-22},
  journaltitle = {Proc. ACM Hum.-Comput. Interact.},
  volume = {5},
  pages = {7:1--7:23},
  doi = {10.1145/3449081},
  issue = {CSCW1}
}

@article{ramdasIdentifyingActionableAlgorithmic2022,
  title = {Identifying an {{Actionable Algorithmic Transparency Framework}}: {{A Comparative Analysis}} of {{Initiatives}} to {{Enhance Accountability}} of {{Social Media Platforms}}},
  author = {Ramdas, Varun},
  date = {2022},
  journaltitle = {Nat'l LU Delhi Stud. LJ},
  volume = {4},
  pages = {74},
  publisher = {HeinOnline}
}

@online{ResponsibleAIKPMG2023,
  title = {Responsible {{AI}}},
  date = {2023-12-15},
  url = {https://kpmg.com/nl/en/home/services/advisory/trusted-enterprise/responsible-ai.html},
  organization = {KPMG}
}

@article{rosenbaumAlgorithmicAccountabilityDigital2019,
  title = {Algorithmic Accountability and Digital Justice: {{A}} Critical Assessment of Technical and Sociotechnical Approaches},
  author = {Rosenbaum, Howard and Fichman, Pnina},
  date = {2019},
  journaltitle = {Proceedings of the Association for Information Science and Technology},
  volume = {56},
  number = {1},
  pages = {237--244},
  publisher = {Wiley Online Library}
}

@inproceedings{sandvigAuditingAlgorithmsResearch2014,
  title = {Auditing Algorithms: {{Research}} Methods for Detecting Discrimination on Internet Platforms},
  booktitle = {Data and {{Discrimination}}: {{Converting Critical Concerns}} into {{Productive}}: {{A}} Preconference at the 64th {{Annual Meeting}} of the {{International Communication Association}}. {{Seattle}}},
  author = {Sandvig, Christian and Hamilton, Kevin and Karahalios, Karrie and Langbort, Cedric},
  date = {2014},
  volume = {22},
  pages = {4349--4357},
  eventtitle = {Data and {{Discrimination}}: {{Converting Critical Concerns}} into {{Productive}}: {{A}} Preconference at the 64th {{Annual Meeting}} of the {{International Communication Association}}. {{Seattle}}}
}

@inproceedings{sculleyHiddenTechnicalDebt2015,
  title = {Hidden Technical Debt in Machine Learning Systems},
  booktitle = {Proceedings of the 28th {{International Conference}} on {{Neural Information Processing Systems}} - {{Volume}} 2},
  author = {Sculley, D. and Holt, Gary and Golovin, Daniel and Davydov, Eugene and Phillips, Todd and Ebner, Dietmar and Chaudhary, Vinay and Young, Michael and Crespo, Jean-Francois and Dennison, Dan},
  date = {2015-12-07},
  series = {{{NIPS}}'15},
  pages = {2503--2511},
  publisher = {MIT Press},
  location = {Cambridge, MA, USA},
  url = {https://proceedings.neurips.cc/paper_files/paper/2015/file/86df7dcfd896fcaf2674f757a2463eba-Paper.pdf}
}

@article{seaverAlgorithmsCultureTactics2017,
  title = {Algorithms as Culture: {{Some}} Tactics for the Ethnography of Algorithmic Systems},
  shorttitle = {Algorithms as Culture},
  author = {Seaver, Nick},
  date = {2017},
  journaltitle = {Big Data \& Society},
  volume = {4},
  number = {2},
  pages = {205395171773810},
  doi = {10/gd8fdx}
}

@legislation{senwydenronAlgorithmicAccountabilityAct2022,
  title = {Algorithmic {{Accountability Act}} 2022},
  editora = {Sen. Wyden, Ron [D-OR]},
  editoratype = {collaborator},
  date = {2022-03-02},
  journaltitle = {Sen. Wyden, Ron [D-OR]}
}

@article{shenEverydayAlgorithmAuditing2021,
  title = {Everyday Algorithm Auditing: {{Understanding}} the Power of Everyday Users in Surfacing Harmful Algorithmic Behaviors},
  author = {Shen, Hong and DeVos, Alicia and Eslami, Motahhare and Holstein, Kenneth},
  date = {2021},
  journaltitle = {Proceedings of the ACM on Human-Computer Interaction},
  volume = {5},
  pages = {1--29},
  publisher = {ACM New York, NY, USA},
  issue = {CSCW2}
}

@inproceedings{smithManyFacesFairness2023,
  title = {The {{Many Faces}} of {{Fairness}}: {{Exploring}} the {{Institutional Logics}} of {{Multistakeholder Microlending Recommendation}}},
  shorttitle = {The {{Many Faces}} of {{Fairness}}},
  booktitle = {Proceedings of the 2023 {{ACM Conference}} on {{Fairness}}, {{Accountability}}, and {{Transparency}}},
  author = {Smith, Jessie J. and Buhayh, Anas and Kathait, Anushka and Ragothaman, Pradeep and Mattei, Nicholas and Burke, Robin and Voida, Amy},
  date = {2023-06-12},
  series = {{{FAccT}} '23},
  pages = {1652--1663},
  publisher = {Association for Computing Machinery},
  location = {New York, NY, USA},
  doi = {10.1145/3593013.3594106},
}

@report{StandardOperatingProcedure2022,
  title = {Standard {{Operating Procedure}} ({{SOP}}) for the Overt Deployment of {{Live Facial Recognition Technology}} ({{LFR}})},
  date = {2022-11-29},
  institution = {Metropolitan Police},
  url = {https://www.met.police.uk/SysSiteAssets/media/downloads/force-content/met/advice/lfr/policy-documents/lfr-sop.pdf}
}

@article{terzisInteroperabilityGovernanceEuropean2023,
  title = {Interoperability and Governance in the {{European Health Data Space Regulation}}},
  author = {Terzis, Petros and Santamaria Echeverria, (Enrique) OE},
  date = {2023},
  journaltitle = {Medical Law International},
  volume = {23},
  number = {4},
  pages = {368--376},
  publisher = {SAGE Publications Ltd},
  doi = {10.1177/09685332231165692}
}

@article{ulloaScalingSearchEngine2022,
  title = {Scaling up Search Engine Audits: {{Practical}} Insights for Algorithm Auditing},
  shorttitle = {Scaling up Search Engine Audits},
  author = {Ulloa, Roberto and Makhortykh, Mykola and Urman, Aleksandra},
  date = {2022-05-02},
  journaltitle = {Journal of Information Science},
  pages = {01655515221093029},
  publisher = {SAGE Publications Ltd},
  doi = {10.1177/01655515221093029}
}

@report{UnderstandingAccuracyBias2022,
  title = {Understanding Accuracy and Bias},
  date = {2022-11-29},
  author = {{Metropolitan Police}},
  url = {https://www.met.police.uk/SysSiteAssets/media/downloads/force-content/met/advice/lfr/other-lfr-documents/lfr-accuracy-and-demographic-differential.pdf}
}

@incollection{vealeAdministrationAlgorithmPublic2019,
  title = {Administration by {{Algorithm}}? {{Public Management}} Meets {{Public Sector Machine Learning}}},
  booktitle = {Algorithmic {{Regulation}}},
  author = {Veale, Michael and Brass, Irina},
  editor = {Yeung, Karen and Lodge, Martin},
  date = {2019},
  pages = {121--149},
  publisher = {Oxford University Press},
  location = {Oxford},
  doi = {10/gfzvz8}
}

@incollection{vealeDeniedDesignDataforthcoming,
  title = {Denied by {{Design}}? {{Data Access Rights}} in {{Encrypted Infrastructures}}},
  booktitle = {Researcher {{Access}} to {{Digital Infrastructures}}},
  author = {Veale, Michael},
  editor = {Ausloos, Jef and family=Souza, given=Siddharth, prefix=de, useprefix=true},
  year = {forthcoming},
  publisher = {Cambridge University Press},
  location = {Cambridge},
  url = {https://doi.org/k5mk},
  urldate = {2023-07-25}
}

@book{wolframWritingRulesEurope2014,
  title = {Writing the {{Rules}} for {{Europe}}},
  author = {Wolfram, Kaiser and Schot, Johan},
  date = {2014},
  publisher = {Palgrave Macmillan}
}

@article{wongSeeingToolkitHow2023,
  title = {Seeing {{Like}} a {{Toolkit}}: {{How Toolkits Envision}} the {{Work}} of {{AI Ethics}}},
  shorttitle = {Seeing {{Like}} a {{Toolkit}}},
  author = {Wong, Richmond Y. and Madaio, Michael A. and Merrill, Nick},
  date = {2023-04-16},
  journaltitle = {Proc. ACM Hum.-Comput. Interact.},
  volume = {7},
  pages = {145:1--145:27},
  doi = {10.1145/3579621},
  issue = {CSCW1}
}

@book{botzemPoliticsAccountingRegulation2012,
  title = {The {{Politics}} of {{Accounting Regulation}}: {{Organizing Transnational Standard Setting}} in {{Financial Reporting}}},
  shorttitle = {The {{Politics}} of {{Accounting Regulation}}},
  author = {Botzem, Sebastian},
  date = {2012},
  publisher = {Edward Elgar Publishing},
  pagetotal = {233}
}

@inproceedings{youngConfrontingPowerCorporate2022,
  title = {Confronting {{Power}} and {{Corporate Capture}} at the {{FAccT Conference}}},
  booktitle = {2022 {{ACM Conference}} on {{Fairness}}, {{Accountability}}, and {{Transparency}}},
  author = {Young, Meg and Katell, Michael and Krafft, P.M.},
  date = {2022-06-21},
  pages = {1375--1386},
  publisher = {ACM},
  location = {Seoul},
  doi = {10.1145/3531146.3533194},
  eventtitle = {{{FAccT}} '22: 2022 {{ACM Conference}} on {{Fairness}}, {{Accountability}}, and {{Transparency}}},
}

\end{document}